\documentclass[preprint, amsmath, amssymb, preprintnumbers, superscriptaddress, floatfix, showpacs, showkeys, aps, prl]{revtex4-1}

\usepackage{graphicx}
\usepackage{float}
\usepackage{color}
\usepackage{latexsym}
\usepackage{setspace}
\usepackage{booktabs}
\usepackage{multirow}
\begin{document}

\title{Giant Intrinsic Spin Hall Effect in W$_3$Ta and other A15 Superconductors}

\author{E. Derunova$^{\dagger}$}
\affiliation{Max Plank Institute for Microstructure Physics, Halle, Germany.}
\author{Y. Sun$^{\dagger}$}
\affiliation{Max Planck Institute for Chemical Physics of Solids, Dresden, Germany.}
\author{C. Felser}
\affiliation{Max Planck Institute for Chemical Physics of Solids, Dresden, Germany.}
\author{S. S. P. Parkin}
\affiliation{Max Plank Institute for Microstructure Physics, Halle, Germany.}
\author{B. Yan}
\affiliation{Department of Condensed Matter Physics, Weizmenn Institute of Science, Rehovot, Israel.}
\author{M. N. Ali}
\affiliation{Max Plank Institute for Microstructure Physics, Halle, Germany.}
\email{maz@berkeley.edu}

\begin{abstract}
\indent{}\textbf{The spin Hall effect (SHE) is the conversion of charge current to spin current, and non-magnetic metals with large SHEs are extremely sought after for spintronic applications, but their rarity has stifled widespread use. Here we predict and explain the large intrinsic SHE in $\beta$-W and the A15 family of superconductors: W$_3$Ta, Ta$_3$Sb, and Cr$_3$Ir having spin hall conductivities (SHC) of -2250, -1400, and 1210 $\frac{\hbar}{e}(\Omega cm)^{-1}$, respectively. Combining concepts from topological physics with the dependence of the SHE on the spin Berry curvature (SBC) of the electronic bands, we propose a simple strategy to rapidly search for materials with large intrinsic SHEs based on the following ideas: high symmetry combined with heavy atoms gives rise to multiple Dirac-like crossings in the electronic structure, without sufficient symmetry protection these crossings gap due to spin orbit coupling (SOC), and gapped Dirac crossings create large spin Berry curvature.} 
\end{abstract}

\maketitle


\indent{}The spin Hall effect has become an important topic in recent years not just from a fundamental physics aspect, but also in regards to near future technological application. This is due to a combination of Moore's law limits on traditional Si-based devices and the concurrent rise of spintronics; creating logic and storage devices based on manipulating both spin and current\cite{ohno2016spintronics}. Spintronics has become the next evolution in computing technology and is already seeing commercial technological adoption. In particular, the study of spin transfer phenomena, where the magnetization of a ferromagnet is manipulated through the transfer of spin angular momentum from a spin current, is considered a promising direction\cite{RevModPhys.80.1517, huai2008spin}. However, the creation of large spin currents, ideally at low power, room temperature, and using cheap materials amenable to facile device fabrication, remains a serious challenge. The three major routes to achieving these criteria are a.) using heterostructures of ferromagnetic metals and nonmagnetic semiconductors, b.) using ferromagnetic semiconductors, or c.) using non-magnetic metals and the SHE\cite{RevModPhys.80.1517}. The direct conversion of charge current to spin current via the SHE is highly appealing for device design since it simplifies device heterostructures and reduces fabrication steps. However, the magnitude of the SHE in non-magnetic metals has been low; simple 4d, and 5d, elements have SHCs calculated to be just a few hundred $\frac{\hbar}{e}(\Omega cm)^{-1}$ with the notable exceptions of Ta (BCC), W (BCC), and Pt (FCC)\cite{PhysRevB.77.165117, morota2011indication}. Recently, Weyl semimetals have been explored as SHE materials due to topologically derived SBC with intrinsic SHC magnitudes calculated to be ~700-800$\frac{\hbar}{e}(\Omega cm)^{-1}$\cite{PhysRevLett.117.146403}.
 
\indent{}The ratio of generated spin current to charge current is defined as the spin hall angle (SHA), and larger SHAs are desired for technological applications. Topological insulators such as Bi$_2$Se$_3$ have been heavily investigated in recent years, and various studies have found SHAs widely ranging from 0.01 - 425, hence their effectiveness as spin Hall materials is debated and large-scale fabrication methods for these materials is still lacking\cite{Mellnik2014, Wang2017, PhysRevLett.114.257202, Fan2014, PhysRevLett.113.196601, PhysRevB.90.094403, Jamali2015}. Pt, and recently $\beta$-W, are the only known sputterable materials to host very large SHAs at room temperature\cite{hao2015beta, demasius2016enhanced, zhang2015role}. Here we explain why $\beta$-W has a giant SHE, based on its symmetry and electronic structure, as well as predict several more sputterable, cheap, and giant SHE materials in the A15 family of superconductors. We also propose a simple and rapid search strategy for finding materials with large SHEs that represents the first application of concepts from topological physics in industrial technology and, with tuning of extrinsic parameters, will dramatically impact today's commercially relevant spin based technologies.


\indent{}In the normal SHE, passing a current through a spin Hall material generates an orthogonal spin current which is polarized perpendicularly to both the charge and spin current directions. There are several mechanisms by which the SHE can be achieved and they can be generally grouped into two categories: extrinsic mechanisms and intrinsic mechanisms. Extrinsic mechanisms refer to the methods by which a spin acquires a transverse velocity from the scattering of electrons due to spin-orbit coupling\cite{6516040, PhysRevLett.104.186403}. Impurity and defect scattering are the most common causes.  In the intrinsic mechanism, on the other hand, the spin current is created in between scattering events rather than during them\cite{RevModPhys.87.1213, 6516040}.  

THe SHC can be calculated using a linear-response approach in the Streda-Kubo formalism \cite{streda, kubo}. In this case, the SHC is split into two parts:\cite{PhysRevB.77.165117, PhysRevLett.100.096401} 

\begin{equation} \label{sigma1}
\sigma^{zI}_{xy}=\frac 1 {2\pi N} \sum_k Tr[\hat{J}^S_x\hat{G}^R\hat{J}^C_y\hat{G}^A]_{\omega=0}.
\end{equation} 

\begin{equation} \label{sigma2}
\sigma^{zII}_{xy}=\frac {-1} {4\pi N} \sum_k  \int^0_{-\infty}Tr[\hat{J}^S_x\frac{\partial\hat{G}^R}{\partial \omega}\hat{J}^C_y\hat{G}^R - \hat{J}^S_x\partial\hat{G}^R\hat{J}^C_y\frac{\hat{G}^R}{\partial \omega}  - <R\leftrightarrow A>].
\end{equation} 

Where $\hat{J}^S_x$ is a $s_z$-spin current operator, $\hat{J}^C_y$ is a charge current operator, $\hat{G}^R,\hat{G}^A$ are retarded and advanced Green functions. In presence of anticrossing bands (i.e. Dirac crossings) $\sigma^{zI}_{xy}=0$ and the main contribution to the SHC comes from the $\sigma^{zII}_{xy}$. When the quasiparticle damping rate is equal to 0 (i.e. pure intrinsic regime), $\sigma^{zII}_{xy}$ is reduced to the following expression:

\begin{equation} \label{sigma2b}
\sigma^{zIIb}_{xy}=\frac {1} {N} \sum_{k,l} f(E^l_k) \Omega^l(k).
\end{equation} 
Where $ f(E^l_k) $ is a Fermi distribution function and $ \Omega^l(k)$ is defined by the expression: 
\begin{equation} \label{BC}
\Omega^l(k)=\sum_{m\ne  l} \frac {2\ Im\{(J^S_x)^{ml}(J^C_y)^{lm}\}}{(E^l_k-E^m_k)^2}.
\end{equation} 

The value of $ \Omega^l(k)$ is referred to as spin Berry curvature which is a specific example of Berry curvature; a much broader concept arising from k-dependence of the wavefunction by the following formula \cite{haldane_berry_2004}: 

\begin{equation} \label{BCCont}
\Omega^{ab}_n(k)=\nabla^a_kA^b_n(k)-\nabla^b_kA^a_n(k).
\end{equation} 
\begin{equation} \label{BPhase}
A^a_n(k)=-i<\psi_n(k)|\nabla^a_k\psi_n(k)>.
\end{equation} 

The wavefunction, $\psi_n(k)$ is heavily influenced by the crystalline symmetries which drive orbital hybridization and thus directly influence the Berry curvature and SBC. Bands which create anticrossings but then also form a hybridization gap with the inclusion of spin orbit coupling (SOC) will give rise to a large SBC. This is because the SBC is opposite for bands on either side of the hybridization gap, but when the E$_F$ lies inside the gap, the oppositely signed contributions are not compensated\cite{6516040}. Since the intrinsic SHE is directly proportional to the integration of the SBC of each occupied band, gapped anticrossings generate large SHEs. This is the well known cause of the large intrinsic SHE in Pt, where the Fermi level lies inside gapped anticrossings near the L and X points in the Brillouin zone (BZ) (see Figure S1) \cite{gradhand2012first, PhysRevB.77.165117, RevModPhys.87.1213, PhysRevLett.100.096401}. When the Berry curvature of all occupied bands is integrated over the entire BZ, Pt naturally has a peak in its SHC vs energy plot at the E$_F$. The mixing of orbital character caused by SOC driven hybridization is also visible in Figure S1. Importantly, the magnitude of the SHE is inversely proportional to the size of the SOC-induced band gap; too large of a gap results in a low SHE. 

\indent{}In order to create a large SHA for use in spintronics, large SHE's are desired, and both extrinsic and intrinsic effects can contribute significantly to the overall magnitude of the generated spin current\cite{6516040}. Ideally, maximizing the SHE will be a combined effort of first picking a material with a large intrinsic effect and then maximizing extrinsic effects through interfaces, doping, defect control, etc. Few materials are known with large intrinsic effects, especially materials which are amenable to large scale thin film fabrication through sputtering. However, a simple strategy to search for or design new SHE materials is to maximize intrinsic SHE by maximizing Berry curvature by having the E$_F$ inside as many small-gapped anticrossings as possible.


\indent{}The recent search for topological insulators (TI's), Dirac, Weyl and Nodal Line semimetals has relied on an understanding of the effects of crystal symmetry on the electronic structure. The symmetry of a crystal structure and the orbitals making up the electronic states when bands are crossed determine which topological state is realized when SOC is considered. Electronic states must be mutually orthogonal in order to not hybridize with each other and open a gap. The various symmetry operations all can create degeneracies in the band structure at specific points or along particular directions\cite{doi:10.7566/JPSJ.87.041001, weng2016topological}. Some of these symmetries, or combinations of them, can create and protect a degeneracy from being gapped by the inclusion of SOC. For example, the C$_{2v}$ point group (spinless case) has four irreducible representations, but the C$_{2v}$ double group (spinful case) only has one irreducible representation. Bands with the same irreducible representation (meaning \textit{not} orthogonal) hybridize, so spinful band crossings with only C$_{2v}$ symmetry gap in the presence of SOC\cite{PhysRevB.91.205128}. On the other hand, the presence of a glide mirror \textit{G}, which has a fraction of a primitive unit vector translation operation \textit{t}, results in $G^{2}=e^{-i\mathbf{k}\cdot \mathbf{t}}$ for Bloch states making the glide eigenvalues equal to $\pm e^{-i\mathbf{k}\cdot \mathbf{t}}$ ($\pm i e^{-i\mathbf{k}\cdot \mathbf{t}}$) for spinless (spinful) systems. So regardless of SOC, the glide mirror, which is a non-symmorphic symmetry operation, yields two distinct eigenvalues and therefore the bands are protected from hybridizing and opening up a gap\cite{PhysRevLett.115.126803, ShuoYingYangSymm}.


\indent{}If one analyzes the crystallographic symmetry, it is possible to determine if a system must have unprotected and protected crossings and even along which k-paths they lay. Recently, there has been an intense effort to use symmetry and group theory to create a complete analysis of all topological classes possible in the 230 space groups\cite{Bradlyn2017, Po2017}. It turns out that protected crossings are \textit{not} rare and unprotected crossings are commonplace. However, unlike the usual aim of material scientists working in the topological field, where the goal has been typically to put \textit{protected} crossings at the Fermi level, here the material scientist's goal, for spin Hall (and anomalous Hall in the case of time-reversal symmetry breaking) purposes, is to put \textit{unprotected} crossings at the Fermi level. Good SHE and AHE materials will have enough symmetry to demand crossings, but not the right symmetries to protect those crossings against SOC, at the E$_F$. This means that many of the materials which were once investigated as potential Dirac/Weyl semimetals but had SOC driven gap openings, or were investigated as potential topological insulators but had additional metallic bands, are worth re-examining for their spin and anomalous Hall effects. Figure 1a shows a simple schematic of a ``failed'' TI band structure which, like Pt, results in a peak in the intrinsic SHC. 



\indent{}$\beta$-W, particularly when doped with small amounts of oxygen (few percent), is also known to have an enormous SHA as large as -0.45 which has been successfully used in spin transfer torque devices\cite{doi:10.1063/1.4753947, demasius2016enhanced}. To the best of our knowledge, no explanation has been given for its large SHC and correspondingly large SHA. However, it can be understood with the concepts outlined above; $\beta$-W has a large \textit{intrinsic} SHE due to several unprotected crossings near the Fermi level resulting in a net large SBC. Figure 1b shows the crystal structure of $\beta$-W, a.k.a W$_3$W, the prototype of the A15 structure type (A$_3$X, space group 223, Pm-3n) which is famous for hosting high critical current superconductors like Nb$_3$Sn which are still the most widely used superconductors in technological applications today\cite{STEWART201528}. The structure has two distinct crystallographic sites (\textit{6c} and \textit{2a}) and can be thought of as a BCC lattice made by the X atom with two A atoms in each face of the cube. The BZ for this system is shown in Figure 1c with several key high symmetry points listed. The electronic structures of $\beta$-W, both without and with SOC, are shown in Figure 1d and Figure 1e, respectively. Shown in red are bands which, without SOC, create several Dirac crossings very close to the Fermi level along the $\Gamma$ - X - M lines, as well as below the E$_F$. These crossings are created by C$_2$-rotation and inversion symmetries: for example, the crossing along $\Gamma$ - X at the $E_F$, is protected by C$_2$ rotations along the (010) and (001) axes coupled with an inversion operation. However, the symmetries protecting these crossings all belong to the C$_{2v}$ point group, which, as described earlier, can create degeneracies without SOC but gap due to SOC. As expected, with the inclusion of SOC in Figure 1e, these bands gap out and the Fermi level lies almost within those gaps. Correspondingly, a broad peak in the SHC versus energy calculation (Figure 1f) straddles the energies where the gapped anticrossings lay. This creates ``hotspots'' of SBC, indicated by the intense red and blue areas, in Figure 1g, in the BZ precisely where the anticrossings were. The maximum SHC actually lies approximately 0.5 eV below the E$_F$ at the intersection of several more gapped crossings and if the E$_F$ were lowered by hole doping, without significantly altering the band dispersion characteristics, it is expected that the SHC could be maximized.



\indent{}Figure 2a shows the the electronic structure without SOC of $\beta$-W broken apart by the crystallographic site and orbital contributions to the bands. The \textit{2a} site contributes almost exclusively to the degeneracy at $\Gamma$ via the t$_{2g}$ orbitals while the \textit{6c} site mixes with orbitals from the \textit{2a} site to create the crossings along $\Gamma$ - X - M both near the E$_F$ and below it. Without SOC, $\beta$-W has highly dispersive linear bands and Dirac crossings akin to a Dirac semimetal. Figures 2b and 2c illustrate the extent of the orbital hybrdization when SOC is included which gaps the Dirac crossings. The changing color of the bands, particularly along the M - X direction below E$_F$, indicates the change of character from d$_{x^2-y^2}$ to the e$_g$ and d$_{z^2}$ orbitals, respectively. This is analogous to what occurs in Pt (Figure S1) where strongly hybridized bands gap its Dirac crossing and result in mixed orbital character and large SBC as well. 


\indent{}The band structure of Ta$_3$Ta; a hypothetical A15 version of Ta where both crystallographic sites are occupied by Ta, is shown in Figure 3a. Due to the similarity of Ta and W, this can be thought of as $\beta$-W with 4 electrons removed, while figures 3b and 3c show the band structures for W$_3$Ta and W$_3$Re, respectively, also in the A15 structure type. However in W$_3$Re, the \textit{2a} site has been replaced with Re (adding 1 electron to $\beta$-W) while in W$_3$Ta it was replaced with Ta (subtracting 1 electron from $\beta$-W). As can be seen from the band structures, the major features, including the Dirac crossings seen in $\beta$-W w/o SOC, are preserved because they are symmetry demanded. Interchanging Ta with W and Re does not fundamentally break the crystalline or band symmetries. Also, Figures 3a and 3b imply that Ta/W site ordering is not critical to manifestation of the Dirac crossings and that even in a disordered thin film, as is expected from sputtered growth, the gapped crossings will persist. W$_3$Ta has shifted the Dirac crossings nearly exactly to the Fermi level and has a maximum calculated SHC of ~2250$\frac{\hbar}{e}(\Omega cm)^{-1}$; one of the highest values for any known compound. Thin films of this cheap material, with expected interfacial spin transparency and conductivities similar to $\beta$-W, should have giant SHAs when coupled in heterostructures with Co/Co$_{40}$Fe$_{40}$B$_{20}$/Permalloy.


\indent{}Another A15 compound, Ta$_3$Sb, also has an intriguing electronic structure, as shown in Figures 4a and 4b. The stoichiometric compound has a maximum SHC at E$_F$ of -1400 $\frac{\hbar}{e}(\Omega cm)^{-1}$ (Figure 4c) as well as an 8-fold degenerate Dirac point nearly at E$_F$ at the R-point\cite{Bradlynaaf5037}. When projected to the 001 face, Ta$_3$Sb houses topologically non-trivial (Figure S2) surface states , shown as the orange bands in Figure 4e connecting the conduction bands and valence bands at X and M. Since this compound is also known to superconduct at 0.7 K\cite{narlikar2014superconductors}, future experimental studies on both the SHE in this material as well the interplay of its topological surface states and superconductivity will be of great interest.


\indent{}The state of the art in searching for large SHE materials has been limited by a lack of a rational design strategy and a difficult candidate screening process. Either materials are experimentally investigated in a combinatorial, serendipity driven approach or from a computation driven approach. The strategy has been to first choose material candidates which are easy to fabricate and contain heavy elements, then calculate electronic structures and Wannier functions before finally calculating the intrinsic SHC using the Kubo formalism. If the final SHC calculation showed a large conductivity, the material is experimentally investigated. This method of screening materials is, however, very time and resource intensive primarily due to the requirement of finding good Wannier functions. Depending on the complexity of the crystal and electronic structures, hundreds of projections need to be attempted before a reasonable one is found and the SHC can be calculated, which is why the search for large SHE materials has been largely dominated by theoretical physicists. However, since we know the ``flag'' feature to look for (gapped Dirac crossings near the $E_F$) and given that Dirac crossings can be generated by crystallographic and orbital symmetries, it is possible to dramatically cut the screening time of SHC materials by a.) choosing material candidates with high symmetry and heavy elements to generate gapped Dirac crossings and b.) simply calculating the electronic structures without and with SOC and comparing them to look for the relevant features. Only for candidates with the right features near E$_F$ does the Wannier projection and SHC need to be calculated. With this simple search strategy, materials scientists, chemists, and experimental physicists who don't have expertise in the details of transport theory, can make significant contributions to the field. 

\indent{}To test this strategy, we expanded our search for large SHE materials to the whole A15 family of materials (many of them are also known to be sputterable for thin film fabrication \cite{narlikar2014superconductors, 1347-4065-32-4R-1759, 1062444, LEHMANN1981145, CHU199731}). Electronic structures without and with SOC were calculated for the rest of the family and only compounds with gap-opened crossings within +/- 2eV of the E$_F$ had their SHC calculated. Fitting the pattern, these compounds have maxima in their SHC at energies commensurate with their gapped anticrossings (see supplementary materials). Several more promising SHE materials were found and their calculated SHCs at E$_F$ are listed in Table 1. Also, alloys of other materials presented here should, like with W$_3$Ta, allow the E$_F$ to be adjusted such that the SHC is maximized. Another example of this would be Ti$_{3-x}$V$_x$Pt, where vanadium doping is used to slightly raise the E$_F$ to a peak in its SHC versus energy plot (Figure S15). 

\indent{}In summary, we have explained why $\beta$-W has a giant SHE; due to its many gapped Dirac crossings resulting in a giant SBC and correspondingly giant intrinsic SHE. We also predicted several more giant SHE compounds, including the cheap alloy W$_3$Ta, which are known to be fabricable in thin film form via sputtering. From understanding the flag feature in the electronic structure and how it can be caused by symmetry, we also proposed a simple and rapid search strategy for finding materials with large SHE that should enable material scientists and chemists to contribute to spintronics and make near-future technological impact. Similar to how the search for topological insulators and Dirac/Weyl materials had an explosion of interest and success due to the approachability of the topic from a variety of different fields, the spin Hall field can also benefit from wide interest. Future work applying the search strategy to other families of compounds such as Heuslers, Perovskites, and various intermetallic structural families will show that, like topological materials, large spin Hall materials are actually quite common and that the high-efficiency generation of spin currents is readily achievable in the near future.

\newpage
\bibliography{Lit}

\begin{thebibliography}{44}%
\makeatletter
\providecommand \@ifxundefined [1]{%
 \@ifx{#1\undefined}
}%
\providecommand \@ifnum [1]{%
 \ifnum #1\expandafter \@firstoftwo
 \else \expandafter \@secondoftwo
 \fi
}%
\providecommand \@ifx [1]{%
 \ifx #1\expandafter \@firstoftwo
 \else \expandafter \@secondoftwo
 \fi
}%
\providecommand \natexlab [1]{#1}%
\providecommand \enquote  [1]{``#1''}%
\providecommand \bibnamefont  [1]{#1}%
\providecommand \bibfnamefont [1]{#1}%
\providecommand \citenamefont [1]{#1}%
\providecommand \href@noop [0]{\@secondoftwo}%
\providecommand \href [0]{\begingroup \@sanitize@url \@href}%
\providecommand \@href[1]{\@@startlink{#1}\@@href}%
\providecommand \@@href[1]{\endgroup#1\@@endlink}%
\providecommand \@sanitize@url [0]{\catcode `\\12\catcode `\$12\catcode
  `\&12\catcode `\#12\catcode `\^12\catcode `\_12\catcode `\%12\relax}%
\providecommand \@@startlink[1]{}%
\providecommand \@@endlink[0]{}%
\providecommand \url  [0]{\begingroup\@sanitize@url \@url }%
\providecommand \@url [1]{\endgroup\@href {#1}{\urlprefix }}%
\providecommand \urlprefix  [0]{URL }%
\providecommand \Eprint [0]{\href }%
\providecommand \doibase [0]{http://dx.doi.org/}%
\providecommand \selectlanguage [0]{\@gobble}%
\providecommand \bibinfo  [0]{\@secondoftwo}%
\providecommand \bibfield  [0]{\@secondoftwo}%
\providecommand \translation [1]{[#1]}%
\providecommand \BibitemOpen [0]{}%
\providecommand \bibitemStop [0]{}%
\providecommand \bibitemNoStop [0]{.\EOS\space}%
\providecommand \EOS [0]{\spacefactor3000\relax}%
\providecommand \BibitemShut  [1]{\csname bibitem#1\endcsname}%
\let\auto@bib@innerbib\@empty
\bibitem [{\citenamefont {Ohno}\ \emph {et~al.}(2016)\citenamefont {Ohno},
  \citenamefont {Stiles},\ and\ \citenamefont {Dieny}}]{ohno2016spintronics}%
  \BibitemOpen
  \bibfield  {author} {\bibinfo {author} {\bibfnamefont {H.}~\bibnamefont
  {Ohno}}, \bibinfo {author} {\bibfnamefont {M.~D.}\ \bibnamefont {Stiles}}, \
  and\ \bibinfo {author} {\bibfnamefont {B.}~\bibnamefont {Dieny}},\
  }\href@noop {} {\bibfield  {journal} {\bibinfo  {journal} {Proceedings of the
  IEEE. Institute of Electrical and Electronics Engineers}\ }\textbf {\bibinfo
  {volume} {104}},\ \bibinfo {pages} {1782} (\bibinfo {year}
  {2016})}\BibitemShut {NoStop}%
\bibitem [{\citenamefont {Fert}(2008)}]{RevModPhys.80.1517}%
  \BibitemOpen
  \bibfield  {author} {\bibinfo {author} {\bibfnamefont {A.}~\bibnamefont
  {Fert}},\ }\href {\doibase 10.1103/RevModPhys.80.1517} {\bibfield  {journal}
  {\bibinfo  {journal} {Rev. Mod. Phys.}\ }\textbf {\bibinfo {volume} {80}},\
  \bibinfo {pages} {1517} (\bibinfo {year} {2008})}\BibitemShut {NoStop}%
\bibitem [{\citenamefont {Huai}(2008)}]{huai2008spin}%
  \BibitemOpen
  \bibfield  {author} {\bibinfo {author} {\bibfnamefont {Y.}~\bibnamefont
  {Huai}},\ }\href@noop {} {\bibfield  {journal} {\bibinfo  {journal} {AAPPS
  bulletin}\ }\textbf {\bibinfo {volume} {18}},\ \bibinfo {pages} {33}
  (\bibinfo {year} {2008})}\BibitemShut {NoStop}%
\bibitem [{\citenamefont {Tanaka}\ \emph {et~al.}(2008)\citenamefont {Tanaka},
  \citenamefont {Kontani}, \citenamefont {Naito}, \citenamefont {Naito},
  \citenamefont {Hirashima}, \citenamefont {Yamada},\ and\ \citenamefont
  {Inoue}}]{PhysRevB.77.165117}%
  \BibitemOpen
  \bibfield  {author} {\bibinfo {author} {\bibfnamefont {T.}~\bibnamefont
  {Tanaka}}, \bibinfo {author} {\bibfnamefont {H.}~\bibnamefont {Kontani}},
  \bibinfo {author} {\bibfnamefont {M.}~\bibnamefont {Naito}}, \bibinfo
  {author} {\bibfnamefont {T.}~\bibnamefont {Naito}}, \bibinfo {author}
  {\bibfnamefont {D.~S.}\ \bibnamefont {Hirashima}}, \bibinfo {author}
  {\bibfnamefont {K.}~\bibnamefont {Yamada}}, \ and\ \bibinfo {author}
  {\bibfnamefont {J.}~\bibnamefont {Inoue}},\ }\href {\doibase
  10.1103/PhysRevB.77.165117} {\bibfield  {journal} {\bibinfo  {journal} {Phys.
  Rev. B}\ }\textbf {\bibinfo {volume} {77}},\ \bibinfo {pages} {165117}
  (\bibinfo {year} {2008})}\BibitemShut {NoStop}%
\bibitem [{\citenamefont {Morota}\ \emph {et~al.}(2011)\citenamefont {Morota},
  \citenamefont {Niimi}, \citenamefont {Ohnishi}, \citenamefont {Wei},
  \citenamefont {Tanaka}, \citenamefont {Kontani}, \citenamefont {Kimura},\
  and\ \citenamefont {Otani}}]{morota2011indication}%
  \BibitemOpen
  \bibfield  {author} {\bibinfo {author} {\bibfnamefont {M.}~\bibnamefont
  {Morota}}, \bibinfo {author} {\bibfnamefont {Y.}~\bibnamefont {Niimi}},
  \bibinfo {author} {\bibfnamefont {K.}~\bibnamefont {Ohnishi}}, \bibinfo
  {author} {\bibfnamefont {D.}~\bibnamefont {Wei}}, \bibinfo {author}
  {\bibfnamefont {T.}~\bibnamefont {Tanaka}}, \bibinfo {author} {\bibfnamefont
  {H.}~\bibnamefont {Kontani}}, \bibinfo {author} {\bibfnamefont
  {T.}~\bibnamefont {Kimura}}, \ and\ \bibinfo {author} {\bibfnamefont
  {Y.}~\bibnamefont {Otani}},\ }\href@noop {} {\bibfield  {journal} {\bibinfo
  {journal} {Physical Review B}\ }\textbf {\bibinfo {volume} {83}},\ \bibinfo
  {pages} {174405} (\bibinfo {year} {2011})}\BibitemShut {NoStop}%
\bibitem [{\citenamefont {Sun}\ \emph {et~al.}(2016)\citenamefont {Sun},
  \citenamefont {Zhang}, \citenamefont {Felser},\ and\ \citenamefont
  {Yan}}]{PhysRevLett.117.146403}%
  \BibitemOpen
  \bibfield  {author} {\bibinfo {author} {\bibfnamefont {Y.}~\bibnamefont
  {Sun}}, \bibinfo {author} {\bibfnamefont {Y.}~\bibnamefont {Zhang}}, \bibinfo
  {author} {\bibfnamefont {C.}~\bibnamefont {Felser}}, \ and\ \bibinfo {author}
  {\bibfnamefont {B.}~\bibnamefont {Yan}},\ }\href {\doibase
  10.1103/PhysRevLett.117.146403} {\bibfield  {journal} {\bibinfo  {journal}
  {Phys. Rev. Lett.}\ }\textbf {\bibinfo {volume} {117}},\ \bibinfo {pages}
  {146403} (\bibinfo {year} {2016})}\BibitemShut {NoStop}%
\bibitem [{\citenamefont {Mellnik}\ \emph {et~al.}(2014)\citenamefont
  {Mellnik}, \citenamefont {Lee}, \citenamefont {Richardella}, \citenamefont
  {Grab}, \citenamefont {Mintun}, \citenamefont {Fischer}, \citenamefont
  {Vaezi}, \citenamefont {Manchon}, \citenamefont {Kim}, \citenamefont
  {Samarth},\ and\ \citenamefont {Ralph}}]{Mellnik2014}%
  \BibitemOpen
  \bibfield  {author} {\bibinfo {author} {\bibfnamefont {A.~R.}\ \bibnamefont
  {Mellnik}}, \bibinfo {author} {\bibfnamefont {J.~S.}\ \bibnamefont {Lee}},
  \bibinfo {author} {\bibfnamefont {A.}~\bibnamefont {Richardella}}, \bibinfo
  {author} {\bibfnamefont {J.~L.}\ \bibnamefont {Grab}}, \bibinfo {author}
  {\bibfnamefont {P.~J.}\ \bibnamefont {Mintun}}, \bibinfo {author}
  {\bibfnamefont {M.~H.}\ \bibnamefont {Fischer}}, \bibinfo {author}
  {\bibfnamefont {A.}~\bibnamefont {Vaezi}}, \bibinfo {author} {\bibfnamefont
  {A.}~\bibnamefont {Manchon}}, \bibinfo {author} {\bibfnamefont {E.-A.}\
  \bibnamefont {Kim}}, \bibinfo {author} {\bibfnamefont {N.}~\bibnamefont
  {Samarth}}, \ and\ \bibinfo {author} {\bibfnamefont {D.~C.}\ \bibnamefont
  {Ralph}},\ }\href {http://dx.doi.org/10.1038/nature13534} {\bibfield
  {journal} {\bibinfo  {journal} {Nature}\ }\textbf {\bibinfo {volume} {511}},\
  \bibinfo {pages} {449 EP } (\bibinfo {year} {2014})}\BibitemShut {NoStop}%
\bibitem [{\citenamefont {Wang}\ \emph {et~al.}(2017)\citenamefont {Wang},
  \citenamefont {Zhu}, \citenamefont {Wu}, \citenamefont {Yang}, \citenamefont
  {Yu}, \citenamefont {Ramaswamy}, \citenamefont {Mishra}, \citenamefont {Shi},
  \citenamefont {Elyasi}, \citenamefont {Teo}, \citenamefont {Wu},\ and\
  \citenamefont {Yang}}]{Wang2017}%
  \BibitemOpen
  \bibfield  {author} {\bibinfo {author} {\bibfnamefont {Y.}~\bibnamefont
  {Wang}}, \bibinfo {author} {\bibfnamefont {D.}~\bibnamefont {Zhu}}, \bibinfo
  {author} {\bibfnamefont {Y.}~\bibnamefont {Wu}}, \bibinfo {author}
  {\bibfnamefont {Y.}~\bibnamefont {Yang}}, \bibinfo {author} {\bibfnamefont
  {J.}~\bibnamefont {Yu}}, \bibinfo {author} {\bibfnamefont {R.}~\bibnamefont
  {Ramaswamy}}, \bibinfo {author} {\bibfnamefont {R.}~\bibnamefont {Mishra}},
  \bibinfo {author} {\bibfnamefont {S.}~\bibnamefont {Shi}}, \bibinfo {author}
  {\bibfnamefont {M.}~\bibnamefont {Elyasi}}, \bibinfo {author} {\bibfnamefont
  {K.-L.}\ \bibnamefont {Teo}}, \bibinfo {author} {\bibfnamefont
  {Y.}~\bibnamefont {Wu}}, \ and\ \bibinfo {author} {\bibfnamefont
  {H.}~\bibnamefont {Yang}},\ }\href {\doibase 10.1038/s41467-017-01583-4}
  {\bibfield  {journal} {\bibinfo  {journal} {Nature Communications}\ }\textbf
  {\bibinfo {volume} {8}},\ \bibinfo {pages} {1364} (\bibinfo {year}
  {2017})}\BibitemShut {NoStop}%
\bibitem [{\citenamefont {Wang}\ \emph {et~al.}(2015)\citenamefont {Wang},
  \citenamefont {Deorani}, \citenamefont {Banerjee}, \citenamefont {Koirala},
  \citenamefont {Brahlek}, \citenamefont {Oh},\ and\ \citenamefont
  {Yang}}]{PhysRevLett.114.257202}%
  \BibitemOpen
  \bibfield  {author} {\bibinfo {author} {\bibfnamefont {Y.}~\bibnamefont
  {Wang}}, \bibinfo {author} {\bibfnamefont {P.}~\bibnamefont {Deorani}},
  \bibinfo {author} {\bibfnamefont {K.}~\bibnamefont {Banerjee}}, \bibinfo
  {author} {\bibfnamefont {N.}~\bibnamefont {Koirala}}, \bibinfo {author}
  {\bibfnamefont {M.}~\bibnamefont {Brahlek}}, \bibinfo {author} {\bibfnamefont
  {S.}~\bibnamefont {Oh}}, \ and\ \bibinfo {author} {\bibfnamefont
  {H.}~\bibnamefont {Yang}},\ }\href {\doibase 10.1103/PhysRevLett.114.257202}
  {\bibfield  {journal} {\bibinfo  {journal} {Phys. Rev. Lett.}\ }\textbf
  {\bibinfo {volume} {114}},\ \bibinfo {pages} {257202} (\bibinfo {year}
  {2015})}\BibitemShut {NoStop}%
\bibitem [{\citenamefont {Fan}\ \emph {et~al.}(2014)\citenamefont {Fan},
  \citenamefont {Upadhyaya}, \citenamefont {Kou}, \citenamefont {Lang},
  \citenamefont {Takei}, \citenamefont {Wang}, \citenamefont {Tang},
  \citenamefont {He}, \citenamefont {Chang}, \citenamefont {Montazeri},
  \citenamefont {Yu}, \citenamefont {Jiang}, \citenamefont {Nie}, \citenamefont
  {Schwartz}, \citenamefont {Tserkovnyak},\ and\ \citenamefont
  {Wang}}]{Fan2014}%
  \BibitemOpen
  \bibfield  {author} {\bibinfo {author} {\bibfnamefont {Y.}~\bibnamefont
  {Fan}}, \bibinfo {author} {\bibfnamefont {P.}~\bibnamefont {Upadhyaya}},
  \bibinfo {author} {\bibfnamefont {X.}~\bibnamefont {Kou}}, \bibinfo {author}
  {\bibfnamefont {M.}~\bibnamefont {Lang}}, \bibinfo {author} {\bibfnamefont
  {S.}~\bibnamefont {Takei}}, \bibinfo {author} {\bibfnamefont
  {Z.}~\bibnamefont {Wang}}, \bibinfo {author} {\bibfnamefont {J.}~\bibnamefont
  {Tang}}, \bibinfo {author} {\bibfnamefont {L.}~\bibnamefont {He}}, \bibinfo
  {author} {\bibfnamefont {L.-T.}\ \bibnamefont {Chang}}, \bibinfo {author}
  {\bibfnamefont {M.}~\bibnamefont {Montazeri}}, \bibinfo {author}
  {\bibfnamefont {G.}~\bibnamefont {Yu}}, \bibinfo {author} {\bibfnamefont
  {W.}~\bibnamefont {Jiang}}, \bibinfo {author} {\bibfnamefont
  {T.}~\bibnamefont {Nie}}, \bibinfo {author} {\bibfnamefont {R.~N.}\
  \bibnamefont {Schwartz}}, \bibinfo {author} {\bibfnamefont {Y.}~\bibnamefont
  {Tserkovnyak}}, \ and\ \bibinfo {author} {\bibfnamefont {K.~L.}\ \bibnamefont
  {Wang}},\ }\href {http://dx.doi.org/10.1038/nmat3973} {\bibfield  {journal}
  {\bibinfo  {journal} {Nature Materials}\ }\textbf {\bibinfo {volume} {13}},\
  \bibinfo {pages} {699 EP } (\bibinfo {year} {2014})},\ \bibinfo {note}
  {article}\BibitemShut {NoStop}%
\bibitem [{\citenamefont {Shiomi}\ \emph {et~al.}(2014)\citenamefont {Shiomi},
  \citenamefont {Nomura}, \citenamefont {Kajiwara}, \citenamefont {Eto},
  \citenamefont {Novak}, \citenamefont {Segawa}, \citenamefont {Ando},\ and\
  \citenamefont {Saitoh}}]{PhysRevLett.113.196601}%
  \BibitemOpen
  \bibfield  {author} {\bibinfo {author} {\bibfnamefont {Y.}~\bibnamefont
  {Shiomi}}, \bibinfo {author} {\bibfnamefont {K.}~\bibnamefont {Nomura}},
  \bibinfo {author} {\bibfnamefont {Y.}~\bibnamefont {Kajiwara}}, \bibinfo
  {author} {\bibfnamefont {K.}~\bibnamefont {Eto}}, \bibinfo {author}
  {\bibfnamefont {M.}~\bibnamefont {Novak}}, \bibinfo {author} {\bibfnamefont
  {K.}~\bibnamefont {Segawa}}, \bibinfo {author} {\bibfnamefont
  {Y.}~\bibnamefont {Ando}}, \ and\ \bibinfo {author} {\bibfnamefont
  {E.}~\bibnamefont {Saitoh}},\ }\href {\doibase
  10.1103/PhysRevLett.113.196601} {\bibfield  {journal} {\bibinfo  {journal}
  {Phys. Rev. Lett.}\ }\textbf {\bibinfo {volume} {113}},\ \bibinfo {pages}
  {196601} (\bibinfo {year} {2014})}\BibitemShut {NoStop}%
\bibitem [{\citenamefont {Deorani}\ \emph {et~al.}(2014)\citenamefont
  {Deorani}, \citenamefont {Son}, \citenamefont {Banerjee}, \citenamefont
  {Koirala}, \citenamefont {Brahlek}, \citenamefont {Oh},\ and\ \citenamefont
  {Yang}}]{PhysRevB.90.094403}%
  \BibitemOpen
  \bibfield  {author} {\bibinfo {author} {\bibfnamefont {P.}~\bibnamefont
  {Deorani}}, \bibinfo {author} {\bibfnamefont {J.}~\bibnamefont {Son}},
  \bibinfo {author} {\bibfnamefont {K.}~\bibnamefont {Banerjee}}, \bibinfo
  {author} {\bibfnamefont {N.}~\bibnamefont {Koirala}}, \bibinfo {author}
  {\bibfnamefont {M.}~\bibnamefont {Brahlek}}, \bibinfo {author} {\bibfnamefont
  {S.}~\bibnamefont {Oh}}, \ and\ \bibinfo {author} {\bibfnamefont
  {H.}~\bibnamefont {Yang}},\ }\href {\doibase 10.1103/PhysRevB.90.094403}
  {\bibfield  {journal} {\bibinfo  {journal} {Phys. Rev. B}\ }\textbf {\bibinfo
  {volume} {90}},\ \bibinfo {pages} {094403} (\bibinfo {year}
  {2014})}\BibitemShut {NoStop}%
\bibitem [{\citenamefont {Jamali}\ \emph {et~al.}(2015)\citenamefont {Jamali},
  \citenamefont {Lee}, \citenamefont {Jeong}, \citenamefont {Mahfouzi},
  \citenamefont {Lv}, \citenamefont {Zhao}, \citenamefont {Nikolic},
  \citenamefont {Mkhoyan}, \citenamefont {Samarth},\ and\ \citenamefont
  {Wang}}]{Jamali2015}%
  \BibitemOpen
  \bibfield  {author} {\bibinfo {author} {\bibfnamefont {M.}~\bibnamefont
  {Jamali}}, \bibinfo {author} {\bibfnamefont {J.~S.}\ \bibnamefont {Lee}},
  \bibinfo {author} {\bibfnamefont {J.~S.}\ \bibnamefont {Jeong}}, \bibinfo
  {author} {\bibfnamefont {F.}~\bibnamefont {Mahfouzi}}, \bibinfo {author}
  {\bibfnamefont {Y.}~\bibnamefont {Lv}}, \bibinfo {author} {\bibfnamefont
  {Z.}~\bibnamefont {Zhao}}, \bibinfo {author} {\bibfnamefont {B.~K.}\
  \bibnamefont {Nikolic}}, \bibinfo {author} {\bibfnamefont {K.~A.}\
  \bibnamefont {Mkhoyan}}, \bibinfo {author} {\bibfnamefont {N.}~\bibnamefont
  {Samarth}}, \ and\ \bibinfo {author} {\bibfnamefont {J.-P.}\ \bibnamefont
  {Wang}},\ }\href {\doibase 10.1021/acs.nanolett.5b03274} {\bibfield
  {journal} {\bibinfo  {journal} {Nano Letters}\ }\textbf {\bibinfo {volume}
  {15}},\ \bibinfo {pages} {7126} (\bibinfo {year} {2015})}\BibitemShut
  {NoStop}%
\bibitem [{\citenamefont {Hao}\ \emph {et~al.}(2015)\citenamefont {Hao},
  \citenamefont {Chen},\ and\ \citenamefont {Xiao}}]{hao2015beta}%
  \BibitemOpen
  \bibfield  {author} {\bibinfo {author} {\bibfnamefont {Q.}~\bibnamefont
  {Hao}}, \bibinfo {author} {\bibfnamefont {W.}~\bibnamefont {Chen}}, \ and\
  \bibinfo {author} {\bibfnamefont {G.}~\bibnamefont {Xiao}},\ }\href@noop {}
  {\bibfield  {journal} {\bibinfo  {journal} {Applied Physics Letters}\
  }\textbf {\bibinfo {volume} {106}},\ \bibinfo {pages} {182403} (\bibinfo
  {year} {2015})}\BibitemShut {NoStop}%
\bibitem [{\citenamefont {Demasius}\ \emph {et~al.}(2016)\citenamefont
  {Demasius}, \citenamefont {Phung}, \citenamefont {Zhang}, \citenamefont
  {Hughes}, \citenamefont {Yang}, \citenamefont {Kellock}, \citenamefont {Han},
  \citenamefont {Pushp},\ and\ \citenamefont {Parkin}}]{demasius2016enhanced}%
  \BibitemOpen
  \bibfield  {author} {\bibinfo {author} {\bibfnamefont {K.-U.}\ \bibnamefont
  {Demasius}}, \bibinfo {author} {\bibfnamefont {T.}~\bibnamefont {Phung}},
  \bibinfo {author} {\bibfnamefont {W.}~\bibnamefont {Zhang}}, \bibinfo
  {author} {\bibfnamefont {B.~P.}\ \bibnamefont {Hughes}}, \bibinfo {author}
  {\bibfnamefont {S.-H.}\ \bibnamefont {Yang}}, \bibinfo {author}
  {\bibfnamefont {A.}~\bibnamefont {Kellock}}, \bibinfo {author} {\bibfnamefont
  {W.}~\bibnamefont {Han}}, \bibinfo {author} {\bibfnamefont {A.}~\bibnamefont
  {Pushp}}, \ and\ \bibinfo {author} {\bibfnamefont {S.~S.}\ \bibnamefont
  {Parkin}},\ }\href@noop {} {\bibfield  {journal} {\bibinfo  {journal} {Nature
  communications}\ }\textbf {\bibinfo {volume} {7}} (\bibinfo {year}
  {2016})}\BibitemShut {NoStop}%
\bibitem [{\citenamefont {Zhang}\ \emph {et~al.}(2015)\citenamefont {Zhang},
  \citenamefont {Han}, \citenamefont {Jiang}, \citenamefont {Yang},\ and\
  \citenamefont {Parkin}}]{zhang2015role}%
  \BibitemOpen
  \bibfield  {author} {\bibinfo {author} {\bibfnamefont {W.}~\bibnamefont
  {Zhang}}, \bibinfo {author} {\bibfnamefont {W.}~\bibnamefont {Han}}, \bibinfo
  {author} {\bibfnamefont {X.}~\bibnamefont {Jiang}}, \bibinfo {author}
  {\bibfnamefont {S.-H.}\ \bibnamefont {Yang}}, \ and\ \bibinfo {author}
  {\bibfnamefont {S.~S.}\ \bibnamefont {Parkin}},\ }\href@noop {} {\bibfield
  {journal} {\bibinfo  {journal} {Nature Physics}\ }\textbf {\bibinfo {volume}
  {11}},\ \bibinfo {pages} {496} (\bibinfo {year} {2015})}\BibitemShut
  {NoStop}%
\bibitem [{\citenamefont {Hoffmann}(2013)}]{6516040}%
  \BibitemOpen
  \bibfield  {author} {\bibinfo {author} {\bibfnamefont {A.}~\bibnamefont
  {Hoffmann}},\ }\href {\doibase 10.1109/TMAG.2013.2262947} {\bibfield
  {journal} {\bibinfo  {journal} {IEEE Transactions on Magnetics}\ }\textbf
  {\bibinfo {volume} {49}},\ \bibinfo {pages} {5172} (\bibinfo {year}
  {2013})}\BibitemShut {NoStop}%
\bibitem [{\citenamefont {Gradhand}\ \emph {et~al.}(2010)\citenamefont
  {Gradhand}, \citenamefont {Fedorov}, \citenamefont {Zahn},\ and\
  \citenamefont {Mertig}}]{PhysRevLett.104.186403}%
  \BibitemOpen
  \bibfield  {author} {\bibinfo {author} {\bibfnamefont {M.}~\bibnamefont
  {Gradhand}}, \bibinfo {author} {\bibfnamefont {D.~V.}\ \bibnamefont
  {Fedorov}}, \bibinfo {author} {\bibfnamefont {P.}~\bibnamefont {Zahn}}, \
  and\ \bibinfo {author} {\bibfnamefont {I.}~\bibnamefont {Mertig}},\ }\href
  {\doibase 10.1103/PhysRevLett.104.186403} {\bibfield  {journal} {\bibinfo
  {journal} {Phys. Rev. Lett.}\ }\textbf {\bibinfo {volume} {104}},\ \bibinfo
  {pages} {186403} (\bibinfo {year} {2010})}\BibitemShut {NoStop}%
\bibitem [{\citenamefont {Sinova}\ \emph
  {et~al.}(2015{\natexlab{a}})\citenamefont {Sinova}, \citenamefont
  {Valenzuela}, \citenamefont {Wunderlich}, \citenamefont {Back},\ and\
  \citenamefont {Jungwirth}}]{RevModPhys.87.1213}%
  \BibitemOpen
  \bibfield  {author} {\bibinfo {author} {\bibfnamefont {J.}~\bibnamefont
  {Sinova}}, \bibinfo {author} {\bibfnamefont {S.~O.}\ \bibnamefont
  {Valenzuela}}, \bibinfo {author} {\bibfnamefont {J.}~\bibnamefont
  {Wunderlich}}, \bibinfo {author} {\bibfnamefont {C.~H.}\ \bibnamefont
  {Back}}, \ and\ \bibinfo {author} {\bibfnamefont {T.}~\bibnamefont
  {Jungwirth}},\ }\href {\doibase 10.1103/RevModPhys.87.1213} {\bibfield
  {journal} {\bibinfo  {journal} {Rev. Mod. Phys.}\ }\textbf {\bibinfo {volume}
  {87}},\ \bibinfo {pages} {1213} (\bibinfo {year}
  {2015}{\natexlab{a}})}\BibitemShut {NoStop}%
\bibitem [{\citenamefont {St\ifmmode~\check{r}\else
  \v{r}\fi{}eda}(2010)}]{streda}%
  \BibitemOpen
  \bibfield  {author} {\bibinfo {author} {\bibfnamefont {P.}~\bibnamefont
  {St\ifmmode~\check{r}\else \v{r}\fi{}eda}},\ }\href {\doibase
  10.1103/PhysRevB.82.045115} {\bibfield  {journal} {\bibinfo  {journal} {Phys.
  Rev. B}\ }\textbf {\bibinfo {volume} {82}},\ \bibinfo {pages} {045115}
  (\bibinfo {year} {2010})}\BibitemShut {NoStop}%
\bibitem [{\citenamefont {Kubo}(1957)}]{kubo}%
  \BibitemOpen
  \bibfield  {author} {\bibinfo {author} {\bibfnamefont {R.}~\bibnamefont
  {Kubo}},\ }\href {\doibase 10.1143/JPSJ.12.570} {\bibfield  {journal}
  {\bibinfo  {journal} {Journal of the Physical Society of Japan}\ }\textbf
  {\bibinfo {volume} {12}},\ \bibinfo {pages} {570} (\bibinfo {year}
  {1957})}\BibitemShut {NoStop}%
\bibitem [{\citenamefont {Guo}\ \emph {et~al.}(2008)\citenamefont {Guo},
  \citenamefont {Murakami}, \citenamefont {Chen},\ and\ \citenamefont
  {Nagaosa}}]{PhysRevLett.100.096401}%
  \BibitemOpen
  \bibfield  {author} {\bibinfo {author} {\bibfnamefont {G.~Y.}\ \bibnamefont
  {Guo}}, \bibinfo {author} {\bibfnamefont {S.}~\bibnamefont {Murakami}},
  \bibinfo {author} {\bibfnamefont {T.-W.}\ \bibnamefont {Chen}}, \ and\
  \bibinfo {author} {\bibfnamefont {N.}~\bibnamefont {Nagaosa}},\ }\href
  {\doibase 10.1103/PhysRevLett.100.096401} {\bibfield  {journal} {\bibinfo
  {journal} {Phys. Rev. Lett.}\ }\textbf {\bibinfo {volume} {100}},\ \bibinfo
  {pages} {096401} (\bibinfo {year} {2008})}\BibitemShut {NoStop}%
\bibitem [{\citenamefont {Haldane}(2004)}]{haldane_berry_2004}%
  \BibitemOpen
  \bibfield  {author} {\bibinfo {author} {\bibfnamefont {F.~D.~M.}\
  \bibnamefont {Haldane}},\ }\href {\doibase 10.1103/PhysRevLett.93.206602}
  {\bibfield  {journal} {\bibinfo  {journal} {Physical Review Letters}\
  }\textbf {\bibinfo {volume} {93}} (\bibinfo {year} {2004}),\
  10.1103/PhysRevLett.93.206602}\BibitemShut {NoStop}%
\bibitem [{\citenamefont {Gradhand}\ \emph {et~al.}(2012)\citenamefont
  {Gradhand}, \citenamefont {Fedorov}, \citenamefont {Pientka}, \citenamefont
  {Zahn}, \citenamefont {Mertig},\ and\ \citenamefont
  {Gy{\"o}rffy}}]{gradhand2012first}%
  \BibitemOpen
  \bibfield  {author} {\bibinfo {author} {\bibfnamefont {M.}~\bibnamefont
  {Gradhand}}, \bibinfo {author} {\bibfnamefont {D.}~\bibnamefont {Fedorov}},
  \bibinfo {author} {\bibfnamefont {F.}~\bibnamefont {Pientka}}, \bibinfo
  {author} {\bibfnamefont {P.}~\bibnamefont {Zahn}}, \bibinfo {author}
  {\bibfnamefont {I.}~\bibnamefont {Mertig}}, \ and\ \bibinfo {author}
  {\bibfnamefont {B.}~\bibnamefont {Gy{\"o}rffy}},\ }\href@noop {} {\bibfield
  {journal} {\bibinfo  {journal} {Journal of Physics: Condensed Matter}\
  }\textbf {\bibinfo {volume} {24}},\ \bibinfo {pages} {213202} (\bibinfo
  {year} {2012})}\BibitemShut {NoStop}%
\bibitem [{\citenamefont {Bernevig}\ \emph {et~al.}(2018)\citenamefont
  {Bernevig}, \citenamefont {Weng}, \citenamefont {Fang},\ and\ \citenamefont
  {Dai}}]{doi:10.7566/JPSJ.87.041001}%
  \BibitemOpen
  \bibfield  {author} {\bibinfo {author} {\bibfnamefont {A.}~\bibnamefont
  {Bernevig}}, \bibinfo {author} {\bibfnamefont {H.}~\bibnamefont {Weng}},
  \bibinfo {author} {\bibfnamefont {Z.}~\bibnamefont {Fang}}, \ and\ \bibinfo
  {author} {\bibfnamefont {X.}~\bibnamefont {Dai}},\ }\href {\doibase
  10.7566/JPSJ.87.041001} {\bibfield  {journal} {\bibinfo  {journal} {Journal
  of the Physical Society of Japan}\ }\textbf {\bibinfo {volume} {87}},\
  \bibinfo {pages} {041001} (\bibinfo {year} {2018})}\BibitemShut {NoStop}%
\bibitem [{\citenamefont {Weng}\ \emph {et~al.}(2016)\citenamefont {Weng},
  \citenamefont {Dai},\ and\ \citenamefont {Fang}}]{weng2016topological}%
  \BibitemOpen
  \bibfield  {author} {\bibinfo {author} {\bibfnamefont {H.}~\bibnamefont
  {Weng}}, \bibinfo {author} {\bibfnamefont {X.}~\bibnamefont {Dai}}, \ and\
  \bibinfo {author} {\bibfnamefont {Z.}~\bibnamefont {Fang}},\ }\href@noop {}
  {\bibfield  {journal} {\bibinfo  {journal} {Journal of Physics: Condensed
  Matter}\ }\textbf {\bibinfo {volume} {28}},\ \bibinfo {pages} {303001}
  (\bibinfo {year} {2016})}\BibitemShut {NoStop}%
\bibitem [{\citenamefont {Gibson}\ \emph {et~al.}(2015)\citenamefont {Gibson},
  \citenamefont {Schoop}, \citenamefont {Muechler}, \citenamefont {Xie},
  \citenamefont {Hirschberger}, \citenamefont {Ong}, \citenamefont {Car},\ and\
  \citenamefont {Cava}}]{PhysRevB.91.205128}%
  \BibitemOpen
  \bibfield  {author} {\bibinfo {author} {\bibfnamefont {Q.~D.}\ \bibnamefont
  {Gibson}}, \bibinfo {author} {\bibfnamefont {L.~M.}\ \bibnamefont {Schoop}},
  \bibinfo {author} {\bibfnamefont {L.}~\bibnamefont {Muechler}}, \bibinfo
  {author} {\bibfnamefont {L.~S.}\ \bibnamefont {Xie}}, \bibinfo {author}
  {\bibfnamefont {M.}~\bibnamefont {Hirschberger}}, \bibinfo {author}
  {\bibfnamefont {N.~P.}\ \bibnamefont {Ong}}, \bibinfo {author} {\bibfnamefont
  {R.}~\bibnamefont {Car}}, \ and\ \bibinfo {author} {\bibfnamefont {R.~J.}\
  \bibnamefont {Cava}},\ }\href {\doibase 10.1103/PhysRevB.91.205128}
  {\bibfield  {journal} {\bibinfo  {journal} {Phys. Rev. B}\ }\textbf {\bibinfo
  {volume} {91}},\ \bibinfo {pages} {205128} (\bibinfo {year}
  {2015})}\BibitemShut {NoStop}%
\bibitem [{\citenamefont {Young}\ and\ \citenamefont
  {Kane}(2015)}]{PhysRevLett.115.126803}%
  \BibitemOpen
  \bibfield  {author} {\bibinfo {author} {\bibfnamefont {S.~M.}\ \bibnamefont
  {Young}}\ and\ \bibinfo {author} {\bibfnamefont {C.~L.}\ \bibnamefont
  {Kane}},\ }\href {\doibase 10.1103/PhysRevLett.115.126803} {\bibfield
  {journal} {\bibinfo  {journal} {Phys. Rev. Lett.}\ }\textbf {\bibinfo
  {volume} {115}},\ \bibinfo {pages} {126803} (\bibinfo {year}
  {2015})}\BibitemShut {NoStop}%
\bibitem [{\citenamefont {Yang}\ \emph {et~al.}(2018)\citenamefont {Yang},
  \citenamefont {Yang}, \citenamefont {Derunova}, \citenamefont {Parkin},
  \citenamefont {Yan},\ and\ \citenamefont {Ali}}]{ShuoYingYangSymm}%
  \BibitemOpen
  \bibfield  {author} {\bibinfo {author} {\bibfnamefont {S.-Y.}\ \bibnamefont
  {Yang}}, \bibinfo {author} {\bibfnamefont {H.}~\bibnamefont {Yang}}, \bibinfo
  {author} {\bibfnamefont {E.}~\bibnamefont {Derunova}}, \bibinfo {author}
  {\bibfnamefont {S.~S.~P.}\ \bibnamefont {Parkin}}, \bibinfo {author}
  {\bibfnamefont {B.}~\bibnamefont {Yan}}, \ and\ \bibinfo {author}
  {\bibfnamefont {M.~N.}\ \bibnamefont {Ali}},\ }\href {\doibase
  10.1080/23746149.2017.1414631} {\bibfield  {journal} {\bibinfo  {journal}
  {Advances in Physics: X}\ }\textbf {\bibinfo {volume} {3}},\ \bibinfo {pages}
  {1414631} (\bibinfo {year} {2018})}\BibitemShut {NoStop}%
\bibitem [{\citenamefont {Bradlyn}\ \emph {et~al.}(2017)\citenamefont
  {Bradlyn}, \citenamefont {Elcoro}, \citenamefont {Cano}, \citenamefont
  {Vergniory}, \citenamefont {Wang}, \citenamefont {Felser}, \citenamefont
  {Aroyo},\ and\ \citenamefont {Bernevig}}]{Bradlyn2017}%
  \BibitemOpen
  \bibfield  {author} {\bibinfo {author} {\bibfnamefont {B.}~\bibnamefont
  {Bradlyn}}, \bibinfo {author} {\bibfnamefont {L.}~\bibnamefont {Elcoro}},
  \bibinfo {author} {\bibfnamefont {J.}~\bibnamefont {Cano}}, \bibinfo {author}
  {\bibfnamefont {M.~G.}\ \bibnamefont {Vergniory}}, \bibinfo {author}
  {\bibfnamefont {Z.}~\bibnamefont {Wang}}, \bibinfo {author} {\bibfnamefont
  {C.}~\bibnamefont {Felser}}, \bibinfo {author} {\bibfnamefont {M.~I.}\
  \bibnamefont {Aroyo}}, \ and\ \bibinfo {author} {\bibfnamefont {B.~A.}\
  \bibnamefont {Bernevig}},\ }\href {http://dx.doi.org/10.1038/nature23268}
  {\bibfield  {journal} {\bibinfo  {journal} {Nature}\ }\textbf {\bibinfo
  {volume} {547}},\ \bibinfo {pages} {298 EP } (\bibinfo {year} {2017})},\
  \bibinfo {note} {article}\BibitemShut {NoStop}%
\bibitem [{\citenamefont {Po}\ \emph {et~al.}(2017)\citenamefont {Po},
  \citenamefont {Vishwanath},\ and\ \citenamefont {Watanabe}}]{Po2017}%
  \BibitemOpen
  \bibfield  {author} {\bibinfo {author} {\bibfnamefont {H.~C.}\ \bibnamefont
  {Po}}, \bibinfo {author} {\bibfnamefont {A.}~\bibnamefont {Vishwanath}}, \
  and\ \bibinfo {author} {\bibfnamefont {H.}~\bibnamefont {Watanabe}},\ }\href
  {\doibase 10.1038/s41467-017-00133-2} {\bibfield  {journal} {\bibinfo
  {journal} {Nature Communications}\ }\textbf {\bibinfo {volume} {8}},\
  \bibinfo {pages} {50} (\bibinfo {year} {2017})}\BibitemShut {NoStop}%
\bibitem [{\citenamefont {Pai}\ \emph {et~al.}(2012)\citenamefont {Pai},
  \citenamefont {Liu}, \citenamefont {Li}, \citenamefont {Tseng}, \citenamefont
  {Ralph},\ and\ \citenamefont {Buhrman}}]{doi:10.1063/1.4753947}%
  \BibitemOpen
  \bibfield  {author} {\bibinfo {author} {\bibfnamefont {C.-F.}\ \bibnamefont
  {Pai}}, \bibinfo {author} {\bibfnamefont {L.}~\bibnamefont {Liu}}, \bibinfo
  {author} {\bibfnamefont {Y.}~\bibnamefont {Li}}, \bibinfo {author}
  {\bibfnamefont {H.~W.}\ \bibnamefont {Tseng}}, \bibinfo {author}
  {\bibfnamefont {D.~C.}\ \bibnamefont {Ralph}}, \ and\ \bibinfo {author}
  {\bibfnamefont {R.~A.}\ \bibnamefont {Buhrman}},\ }\href {\doibase
  10.1063/1.4753947} {\bibfield  {journal} {\bibinfo  {journal} {Applied
  Physics Letters}\ }\textbf {\bibinfo {volume} {101}},\ \bibinfo {pages}
  {122404} (\bibinfo {year} {2012})}\BibitemShut {NoStop}%
\bibitem [{\citenamefont {Stewart}(2015)}]{STEWART201528}%
  \BibitemOpen
  \bibfield  {author} {\bibinfo {author} {\bibfnamefont {G.}~\bibnamefont
  {Stewart}},\ }\href {\doibase https://doi.org/10.1016/j.physc.2015.02.013}
  {\bibfield  {journal} {\bibinfo  {journal} {Physica C: Superconductivity and
  its Applications}\ }\textbf {\bibinfo {volume} {514}},\ \bibinfo {pages} {28
  } (\bibinfo {year} {2015})}\BibitemShut {NoStop}%
\bibitem [{\citenamefont {Bradlyn}\ \emph {et~al.}(2016)\citenamefont
  {Bradlyn}, \citenamefont {Cano}, \citenamefont {Wang}, \citenamefont
  {Vergniory}, \citenamefont {Felser}, \citenamefont {Cava},\ and\
  \citenamefont {Bernevig}}]{Bradlynaaf5037}%
  \BibitemOpen
  \bibfield  {author} {\bibinfo {author} {\bibfnamefont {B.}~\bibnamefont
  {Bradlyn}}, \bibinfo {author} {\bibfnamefont {J.}~\bibnamefont {Cano}},
  \bibinfo {author} {\bibfnamefont {Z.}~\bibnamefont {Wang}}, \bibinfo {author}
  {\bibfnamefont {M.~G.}\ \bibnamefont {Vergniory}}, \bibinfo {author}
  {\bibfnamefont {C.}~\bibnamefont {Felser}}, \bibinfo {author} {\bibfnamefont
  {R.~J.}\ \bibnamefont {Cava}}, \ and\ \bibinfo {author} {\bibfnamefont
  {B.~A.}\ \bibnamefont {Bernevig}},\ }\href {\doibase 10.1126/science.aaf5037}
  {\bibfield  {journal} {\bibinfo  {journal} {Science}\ }\textbf {\bibinfo
  {volume} {353}} (\bibinfo {year} {2016}),\
  10.1126/science.aaf5037}\BibitemShut {NoStop}%
\bibitem [{\citenamefont {Narlikar}(2014)}]{narlikar2014superconductors}%
  \BibitemOpen
  \bibfield  {author} {\bibinfo {author} {\bibfnamefont {A.}~\bibnamefont
  {Narlikar}},\ }\href {https://books.google.de/books?id=3iYUDAAAQBAJ} {\emph
  {\bibinfo {title} {Superconductors}}}\ (\bibinfo  {publisher} {OUP Oxford},\
  \bibinfo {year} {2014})\BibitemShut {NoStop}%
\bibitem [{\citenamefont {Arita}\ and\ \citenamefont
  {Nishida}(1993)}]{1347-4065-32-4R-1759}%
  \BibitemOpen
  \bibfield  {author} {\bibinfo {author} {\bibfnamefont {M.}~\bibnamefont
  {Arita}}\ and\ \bibinfo {author} {\bibfnamefont {I.}~\bibnamefont
  {Nishida}},\ }\href {http://stacks.iop.org/1347-4065/32/i=4R/a=1759}
  {\bibfield  {journal} {\bibinfo  {journal} {Japanese Journal of Applied
  Physics}\ }\textbf {\bibinfo {volume} {32}},\ \bibinfo {pages} {1759}
  (\bibinfo {year} {1993})}\BibitemShut {NoStop}%
\bibitem [{\citenamefont {Allen}\ \emph {et~al.}(1983)\citenamefont {Allen},
  \citenamefont {Beasley}, \citenamefont {Hammond},\ and\ \citenamefont
  {Turneaure}}]{1062444}%
  \BibitemOpen
  \bibfield  {author} {\bibinfo {author} {\bibfnamefont {L.}~\bibnamefont
  {Allen}}, \bibinfo {author} {\bibfnamefont {M.}~\bibnamefont {Beasley}},
  \bibinfo {author} {\bibfnamefont {R.}~\bibnamefont {Hammond}}, \ and\
  \bibinfo {author} {\bibfnamefont {J.}~\bibnamefont {Turneaure}},\ }\href
  {\doibase 10.1109/TMAG.1983.1062444} {\bibfield  {journal} {\bibinfo
  {journal} {IEEE Transactions on Magnetics}\ }\textbf {\bibinfo {volume}
  {19}},\ \bibinfo {pages} {1003} (\bibinfo {year} {1983})}\BibitemShut
  {NoStop}%
\bibitem [{\citenamefont {Lehmann}\ \emph {et~al.}(1981)\citenamefont
  {Lehmann}, \citenamefont {Adrian}, \citenamefont {Bieger}, \citenamefont
  {Müller}, \citenamefont {Nölscher}, \citenamefont {Saemann-Ischenko},\ and\
  \citenamefont {Haase}}]{LEHMANN1981145}%
  \BibitemOpen
  \bibfield  {author} {\bibinfo {author} {\bibfnamefont {M.}~\bibnamefont
  {Lehmann}}, \bibinfo {author} {\bibfnamefont {H.}~\bibnamefont {Adrian}},
  \bibinfo {author} {\bibfnamefont {J.}~\bibnamefont {Bieger}}, \bibinfo
  {author} {\bibfnamefont {P.}~\bibnamefont {Müller}}, \bibinfo {author}
  {\bibfnamefont {C.}~\bibnamefont {Nölscher}}, \bibinfo {author}
  {\bibfnamefont {G.}~\bibnamefont {Saemann-Ischenko}}, \ and\ \bibinfo
  {author} {\bibfnamefont {E.}~\bibnamefont {Haase}},\ }\href {\doibase
  https://doi.org/10.1016/0038-1098(81)91066-8} {\bibfield  {journal} {\bibinfo
   {journal} {Solid State Communications}\ }\textbf {\bibinfo {volume} {39}},\
  \bibinfo {pages} {145 } (\bibinfo {year} {1981})}\BibitemShut {NoStop}%
\bibitem [{\citenamefont {Chu}\ \emph {et~al.}(1997)\citenamefont {Chu},
  \citenamefont {Chang},\ and\ \citenamefont {Lee}}]{CHU199731}%
  \BibitemOpen
  \bibfield  {author} {\bibinfo {author} {\bibfnamefont {J.}~\bibnamefont
  {Chu}}, \bibinfo {author} {\bibfnamefont {J.}~\bibnamefont {Chang}}, \ and\
  \bibinfo {author} {\bibfnamefont {P.}~\bibnamefont {Lee}},\ }\href {\doibase
  https://doi.org/10.1016/S0254-0584(97)80180-0} {\bibfield  {journal}
  {\bibinfo  {journal} {Materials Chemistry and Physics}\ }\textbf {\bibinfo
  {volume} {50}},\ \bibinfo {pages} {31 } (\bibinfo {year} {1997})}\BibitemShut
  {NoStop}%
\bibitem [{\citenamefont {Koepernik}\ and\ \citenamefont
  {Eschrig}(1999)}]{Koepernik1999}%
  \BibitemOpen
  \bibfield  {author} {\bibinfo {author} {\bibfnamefont {K.}~\bibnamefont
  {Koepernik}}\ and\ \bibinfo {author} {\bibfnamefont {H.}~\bibnamefont
  {Eschrig}},\ }\href@noop {} {\bibfield  {journal} {\bibinfo  {journal} {Phys.
  Rev. B}\ }\textbf {\bibinfo {volume} {59}},\ \bibinfo {pages} {1743}
  (\bibinfo {year} {1999})}\BibitemShut {NoStop}%
\bibitem [{\citenamefont {Perdew}\ \emph {et~al.}(1996)\citenamefont {Perdew},
  \citenamefont {Burke},\ and\ \citenamefont {Ernzerhof}}]{perdew1996}%
  \BibitemOpen
  \bibfield  {author} {\bibinfo {author} {\bibfnamefont {J.~P.}\ \bibnamefont
  {Perdew}}, \bibinfo {author} {\bibfnamefont {K.}~\bibnamefont {Burke}}, \
  and\ \bibinfo {author} {\bibfnamefont {M.}~\bibnamefont {Ernzerhof}},\
  }\href@noop {} {\bibfield  {journal} {\bibinfo  {journal} {Phys. Rev. Lett.}\
  }\textbf {\bibinfo {volume} {77}},\ \bibinfo {pages} {3865} (\bibinfo {year}
  {1996})}\BibitemShut {NoStop}%
\bibitem [{\citenamefont {Blaha}\ \emph {et~al.}(2001)\citenamefont {Blaha},
  \citenamefont {Schwarz}, \citenamefont {Madsen}, \citenamefont {Kvasnicka},\
  and\ \citenamefont {Luitz}}]{blaha2001}%
  \BibitemOpen
  \bibfield  {author} {\bibinfo {author} {\bibfnamefont {P.}~\bibnamefont
  {Blaha}}, \bibinfo {author} {\bibfnamefont {K.}~\bibnamefont {Schwarz}},
  \bibinfo {author} {\bibfnamefont {G.}~\bibnamefont {Madsen}}, \bibinfo
  {author} {\bibfnamefont {D.}~\bibnamefont {Kvasnicka}}, \ and\ \bibinfo
  {author} {\bibfnamefont {J.}~\bibnamefont {Luitz}},\ }\href@noop {}
  {\bibfield  {journal} {\bibinfo  {journal} {WIEN2k, An Augmented Plane Wave+
  Local Orbitals Program for calculating Crystal Properties, Technische
  Universit{\"a}t Wien, Austria}\ } (\bibinfo {year} {2001})}\BibitemShut
  {NoStop}%
\bibitem [{\citenamefont {Sinova}\ \emph
  {et~al.}(2015{\natexlab{b}})\citenamefont {Sinova}, \citenamefont
  {Valenzuela}, \citenamefont {Wunderlich}, \citenamefont {Back},\ and\
  \citenamefont {Jungwirth}}]{Sinova2015}%
  \BibitemOpen
  \bibfield  {author} {\bibinfo {author} {\bibfnamefont {J.}~\bibnamefont
  {Sinova}}, \bibinfo {author} {\bibfnamefont {S.~O.}\ \bibnamefont
  {Valenzuela}}, \bibinfo {author} {\bibfnamefont {J.}~\bibnamefont
  {Wunderlich}}, \bibinfo {author} {\bibfnamefont {C.}~\bibnamefont {Back}}, \
  and\ \bibinfo {author} {\bibfnamefont {T.}~\bibnamefont {Jungwirth}},\
  }\href@noop {} {\bibfield  {journal} {\bibinfo  {journal} {Rev. Mod. Phys.}\
  }\textbf {\bibinfo {volume} {87}},\ \bibinfo {pages} {1213} (\bibinfo {year}
  {2015}{\natexlab{b}})}\BibitemShut {NoStop}%
\bibitem [{\citenamefont {Xiao}\ \emph {et~al.}(2010)\citenamefont {Xiao},
  \citenamefont {Chang},\ and\ \citenamefont {Niu}}]{Xiao2010}%
  \BibitemOpen
  \bibfield  {author} {\bibinfo {author} {\bibfnamefont {D.}~\bibnamefont
  {Xiao}}, \bibinfo {author} {\bibfnamefont {M.-C.}\ \bibnamefont {Chang}}, \
  and\ \bibinfo {author} {\bibfnamefont {Q.}~\bibnamefont {Niu}},\ }\href@noop
  {} {\bibfield  {journal} {\bibinfo  {journal} {Rev. Mod. Phys.}\ }\textbf
  {\bibinfo {volume} {82}},\ \bibinfo {pages} {1959} (\bibinfo {year}
  {2010})}\BibitemShut {NoStop}%
\end{thebibliography}%
\newpage


\begin{figure}[h]
	\includegraphics[width=1.0\textwidth]{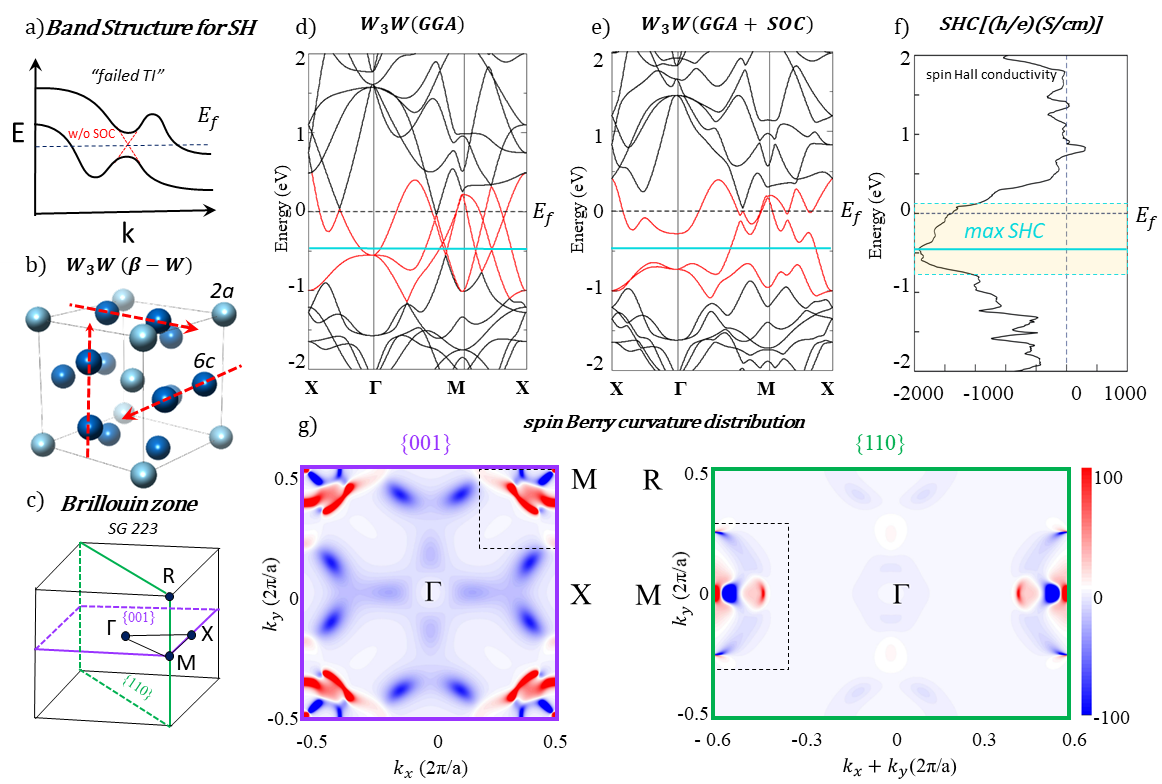}
	\caption{\scriptsize{\textbf{(color online):} Panel \textbf{a}) Schematic of the ``failed TI'' bandstructure which is good for spin Hall purposes. Panel \textbf{b}) Crystal structure of W$_3$W, a.k.a $\beta$-W, prototype of the A15 family. Light and dark blue balls represent the two different crystallographic sites and red arrows show the orthogonal infinite chains formed by atoms at the \textit{6c} site. Panel \textbf{c}) Brillouin Zone (BZ) for A15 family (SG 223). Panel \textbf{d,e}) Electronic structures of W$_3$W without and with SOC, respectively. Dirac crossings are visible (without SOC) along the $\Gamma$-X-M lines, both at and below the E$_F$. Panel \textbf{f}) spin Hall conductivity versus energy plot of W$_3$. Panel \textbf{g}) spin Berry curvature distribution in the the {001} and {110} planes of the BZ of W$_3$W where red and blue areas represent positive and negative regions.}}
	\label{Figure_1}
\end{figure}
\newpage


\begin{figure}[h]
	\includegraphics[width=1.0\textwidth]{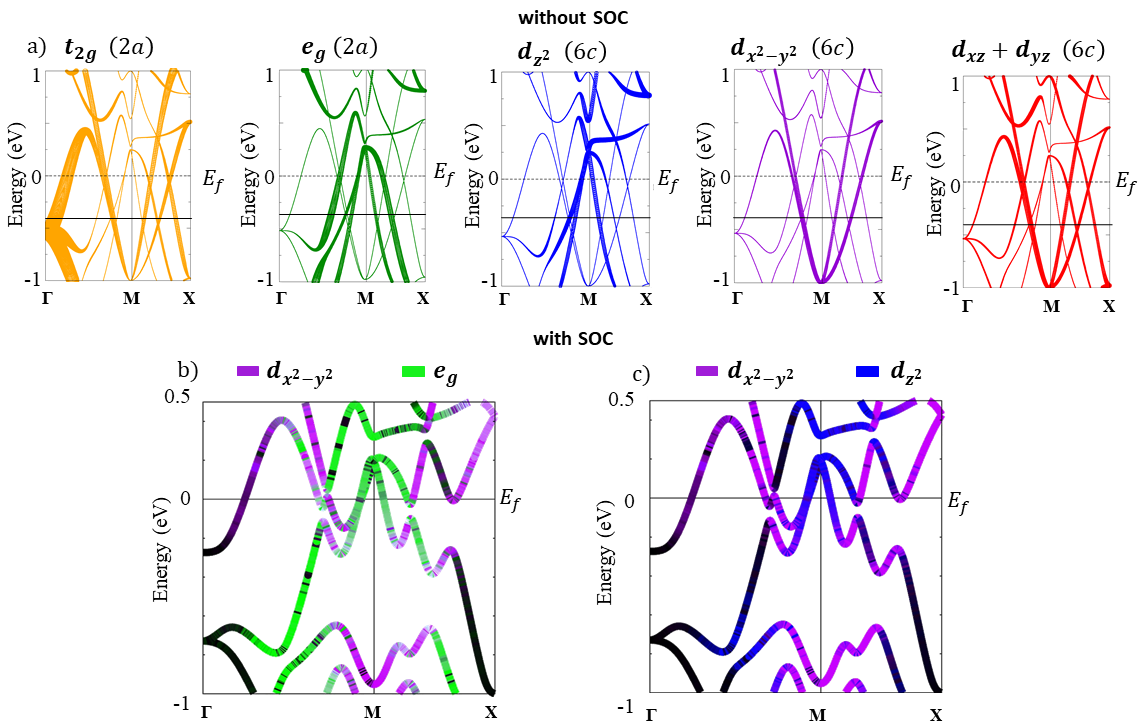}
 \caption{\scriptsize{\textbf{(color online):} Panel \textbf{a}) The electronic structure of W$_3$W near the Dirac crossings decomposed into the various orbital and crystallographic site (\textit{2a} and \textit{6c}) contributions. Thickness of the bands indicates the extent of the orbital or orbital group contribution to a band. By symmetry, the \textit{2a} site's orbitals group into the t$_{2g}$ and e$_g$ sets. The \textit{6c} site has lower degeneracy. The dxy is not shown as it does contribute to the relevant bands. Panel \textbf{b,c}) The electronic structures with SOC included, illustrating the orbital hybridization driven by SOC, opening gaps and generating spin Berry curvature.}}
	\label{Figure_2}
\end{figure}
\newpage


\begin{figure}[h]
	\includegraphics[width=1.0\textwidth]{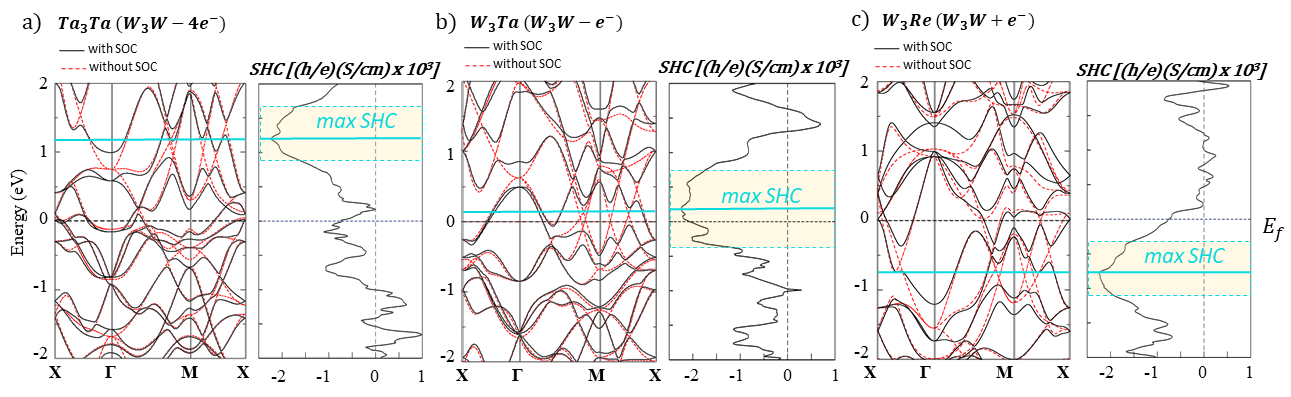}
	\caption{\scriptsize{\textbf{(color online):} Panel \textbf{a,b,c}) Electronic structures of Ta$_3$Ta, W$_3$Re and W$_3$Ta, without and with SOC included, as well as their spin Hall conductivity versus energy plots.}}
	\label{Figure_3}
\end{figure}
\newpage


\begin{figure}[h]
	\includegraphics[width=0.8\textwidth]{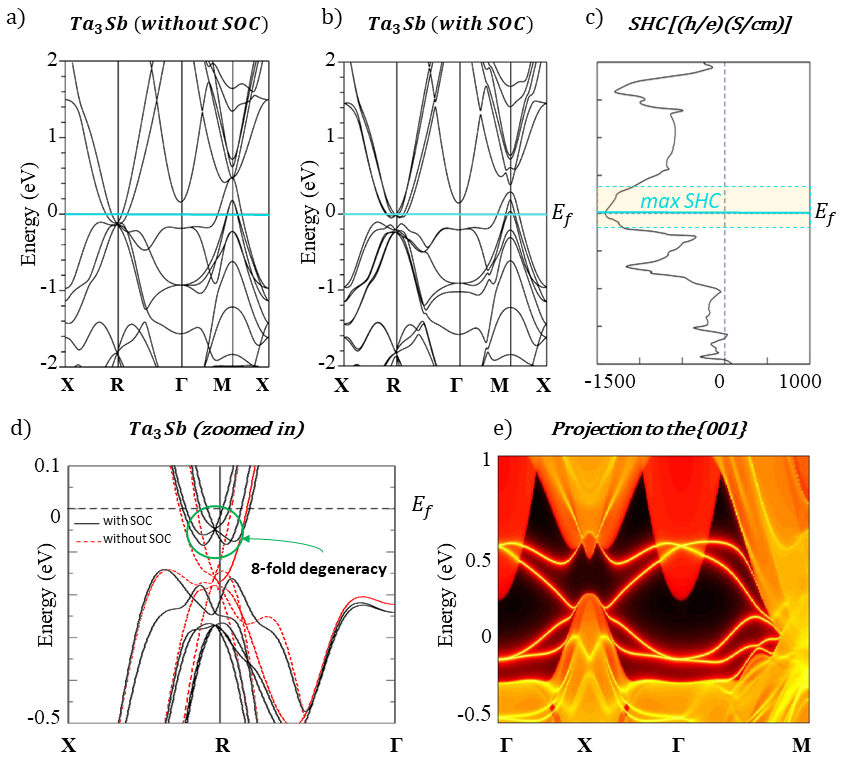}
	\caption{\scriptsize{\textbf{(color online):} Panel \textbf{a,b,c}) Electronic structure of Ta$_3$Sb without and with SOC as well as its SHC versus energy plot. Ta$_3$Sb's has a peak in its SHC at its E$_F$. Panel \textbf{d}) Zoomed in band structure highlighting the 8-fold degeneracy near the E$_F$. Panel \textbf{d} Topological protected surface states (projected to the 001 face) in Ta$_3$Sb connecting the conduction and valence bands along X-$\Gamma$-M.}}
	\label{Figure_4}
\end{figure}


\begin{table}[H]
\caption{\label{tab:SHC}Spin Hall conductivities of selected A15 materials}
\begin{center}
\begin{tabular}{|c|c|c|c|c|c|c|c|c|c|c|c|c|}
\hline
Compounds & W$_3$Ta & Ta$_3$Sb & Cr$_3$Ir & Nb$_3$Au & Ta$_3$Au & W$_3$Re& Nb$_3$Bi & W$_3$Si & Ta$_3$Sn & Nb$_3$Os\\
\hline
SHC ($\frac{\hbar}{e}(\Omega cm)^{-1}$) & -2250 & -1400 & 1209 & -1060 & -870 & -780 & -670 & -640 & -620 & -460\\
\hline
\end{tabular}
\end{center}
\end{table}

\newpage

\textbf{Methods}

\indent{}Our calculations have been performed by using the density-functional theory (DFT) with localized atomic orbital basis and the full potential as implemented in the code of full-potential local-orbital (FPLO)~\cite{Koepernik1999}. The exchange and correlation energy was considered in the generalized gradient approximation (GGA) level~\cite{perdew1996}. The electronic band structures were further confirmed by the calculations from $ab-initio$ code of \textsc{wien2k}~\cite{blaha2001}. In all the calculations we have adopted the experimentally measured lattice structures. By projecting the Bloch wave functions to the high symmetry atomic orbital like Wannier functions, we have constructed the tight binding model Hamiltonian. The intrinsic SHCs were calculated from the model Hamiltonian by the Kubo formula approach in the clean limit ~\cite{Sinova2015, Xiao2010}.

\begin{equation}
\begin{aligned}
\sigma_{ij}^{k}=e\hbar\int_{_{BZ}}\frac{d\vec{k}}{(2\pi)^{3}}\underset{n}{\sum}f_{n\vec{k}}\Omega_{n,ij}^{k}(\vec{k}), \\
\Omega_{n,ij}^{k}(\vec{k})=-2Im\underset{n'\neq n}{\sum}\frac{\langle n\vec{k}| J_{i}^{k}|n'\vec{k} \rangle \langle n'\vec{k}| v_{j}|n\vec{k}\rangle}{(E_{n\vec{k}}-E_{n'\vec{k}})^{2}}
\end{aligned}
\label{SHC}
\end{equation}
The SHC $\sigma_{ij}^{k}$ refers to the spin current ($j_i^{s,k}$) flowing along the $i$-th direction with the spin polarization along $k$, generated by an electric field ($E_j$) along the $j$-th direction, $j_i^{s,k} = \sigma_{ij}^{k} E_j$. The spin current operator is $J_{i}^{k}=\frac{1}{2}\left\{ \begin{array}{cc}{v_{i}}, & {s_{k}}\end{array}\right\}$, with spin operator ${s}$ and velocity operator ${v_{i}}=\frac{1}{\hbar}\frac{\partial {H}}{\partial k_{i}}$ ($i,j,k=x,y,z$). $| n\vec{k} \rangle$ is the $n-th$ eigenvector for the Hamiltonian ${H}$ with eigenvalue $E_{n\vec{k}}$, and $f_{n\vec{k}}$ is the Fermi--Dirac distribution for the $n$-th band. For convenience, we call $\Omega_{n, ij}^{k}(\vec{k})$ as the spin Berry curvature as analogy to the ordinary Berry curvature, and the SHCs were computed by the integral of spin Berry curvature in the BZ with a $500 \times 500 \times 500$ $k$-grid.

\textbf{Acknowledgments}

\indent{}This research was supported by the MPI for Microstructure Physics in Halle, the MPI for Chemical Physics of Solids in Dresden, the Alexander von Humboldt Foundation and their Sofia Kovalevskaja Award, the German Federal Ministry of Education and Research as well as the Minerva Foundation. Also, we acknowledge support by the Ruth and Herman Albert Scholars Program for New Scientists in Weizmann Institute of Science, Israel as well as the German-Israeli Foundation for Scientific Research and Development. The authors also wish to acknowledge James Taylor, Jiho Yoon and Hao Yang for useful discussions.

\textbf{Author Contributions}

\indent{}E.D. and Y.S. are the co-lead authors of this study. They carried out DFT and SHC calculation and analysis. C.F. and S.S.P.P. supported the project. B. Y. and M. N. A. are the primary investigators and directed the research. M.N.A. conceived the project.
 
\textbf{Competing Interests:} The authors declare that they have no competing financial interests.

\textbf{Correspondence:} Correspondence and requests for materials should be addressed to Mazhar N. Ali (email: maz@berkeley.edu).
\newpage

\begin{center}\textbf{Supplementary Information}\end{center}


\section{Pt band structure and spin Hall conductivity}

\begin{figure}[h]
	\includegraphics[width=1.0\textwidth]{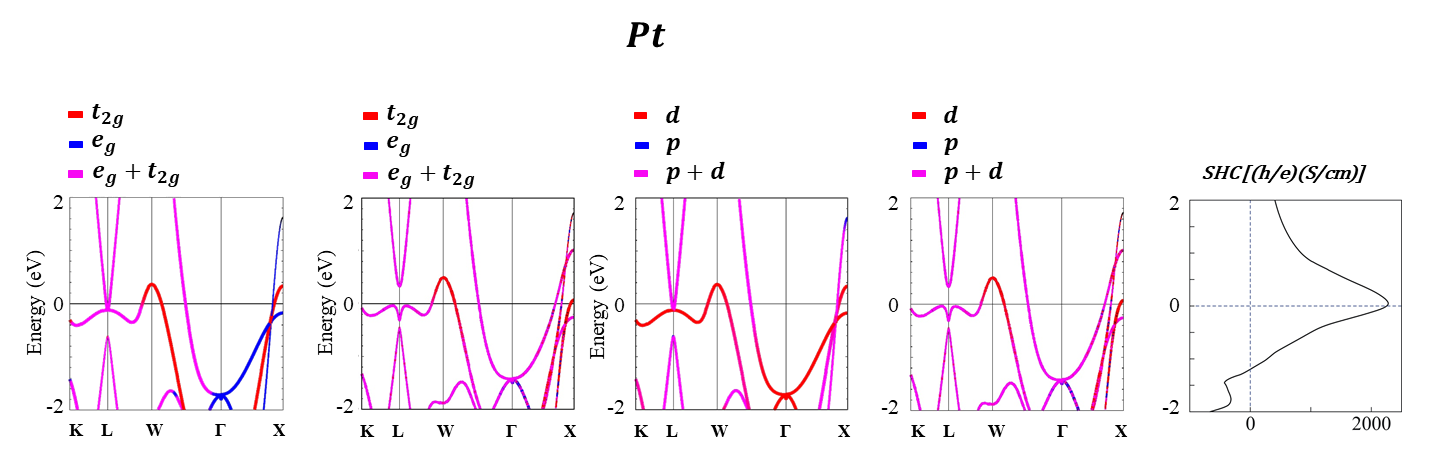}
	\caption{\scriptsize{\textbf{(color online):} From left to right: orbital contribution decomposition of the electronic band structure of platinum followed by the calculated spin Hall conductivity. Gap opened crossings are seen at the L-point and along the $\Gamma$ - X line.}}
	\label{Figure_S1}
\end{figure}
\newpage


\section{Ta$_3$Sb Z$_2$ index}

\begin{figure}[h]
	\includegraphics[width=1\textwidth]{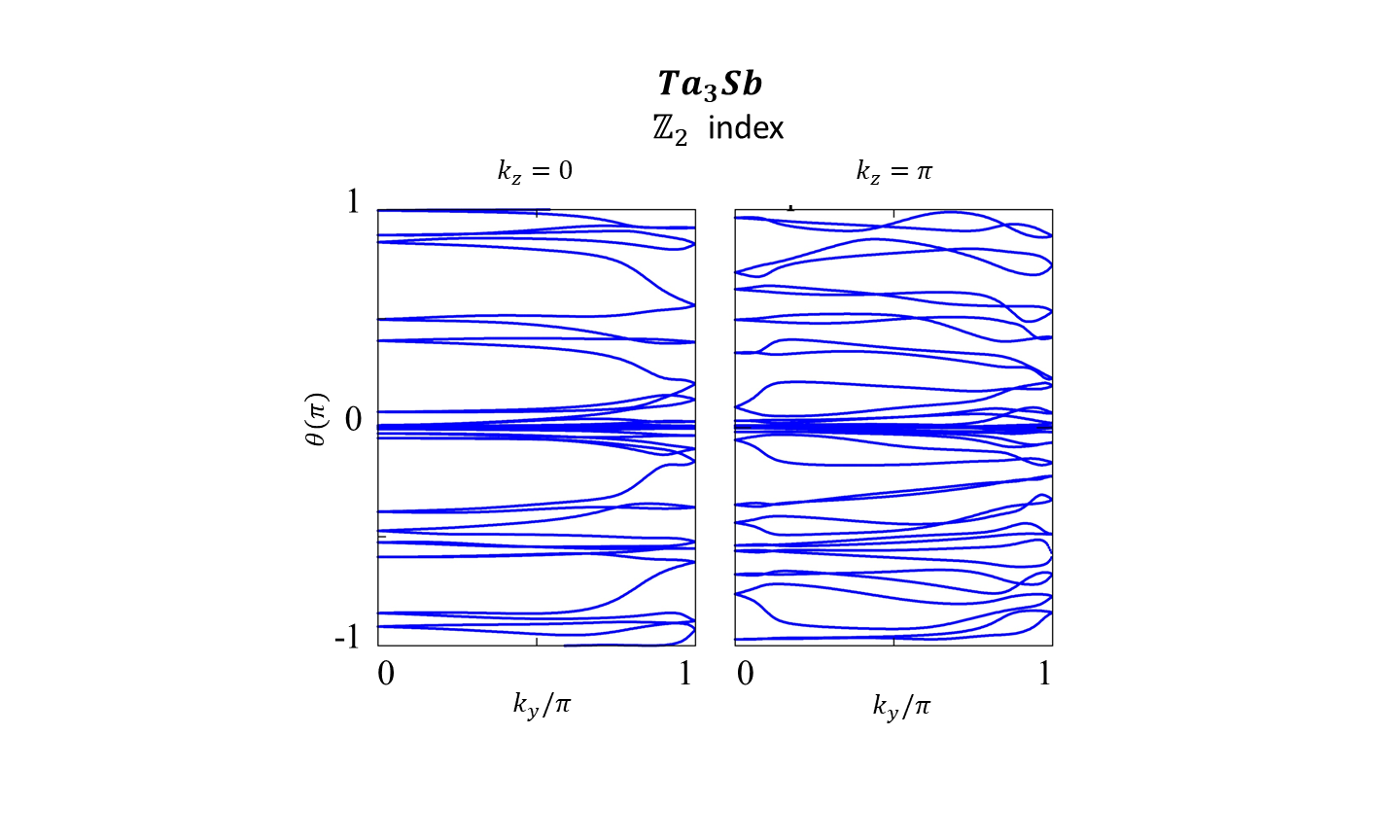}
	\caption{\scriptsize{\textbf{(color online):} Z$_2$ index plot for the Ta$_3$Sb surface states, showing their non-trivial topological nature.}}
	\label{Figure_S2}
\end{figure}
\newpage


\section{A15 band structures and spin Hall conductivities}


\begin{table}[H]
\caption{spin Hall conductivities of calculated A15 materials at E$_F$}
\begin{center}
\begin{tabular}{|c|c|c|c|c|c|c|c|c|c|c|}
\hline
Compounds & W$_3$Ta & W$_3$W & Ta$_3$Sb & Cr$_3$Ir & Nb$_3$Au & Ta$_3$Au & W$_3$Re & Ta$_3$Ta & Nb$_3$Bi & W$_3$Si\\
\hline
SHC ($\frac{\hbar}{e}(\Omega cm)^{-1}$) & -2250 & -1900 & -1400 & 1209 & -1060 & -870 & -780 & -720 & -670 & -640\\
\hline
\end{tabular}
\end{center}
\end{table}


\begin{table}[H]
\caption{spin Hall conductivities of calculated A15 materials at E$_F$ (cont.)}
\begin{center}
\begin{tabular}{|c|c|c|c|c|c|c|c|c|c|}
\hline
Compounds & Ta$_3$Sn & Nb$_3$Os & Nb$_3$Al & V$_3$Pt & Ti$_3$Pt & Ta$_3$Os & Ta$_3$Ir & Ti$_3$Ir & Cr$_3$Os\\
\hline
SHC ($\frac{\hbar}{e}(\Omega cm)^{-1}$) & -620 & -460 & -440 & -440 & 330 & -230 & -143 & 56 & -40\\
\hline
\end{tabular}
\end{center}
\end{table}


\begin{figure}[ht]
	\includegraphics[width=1\textwidth]{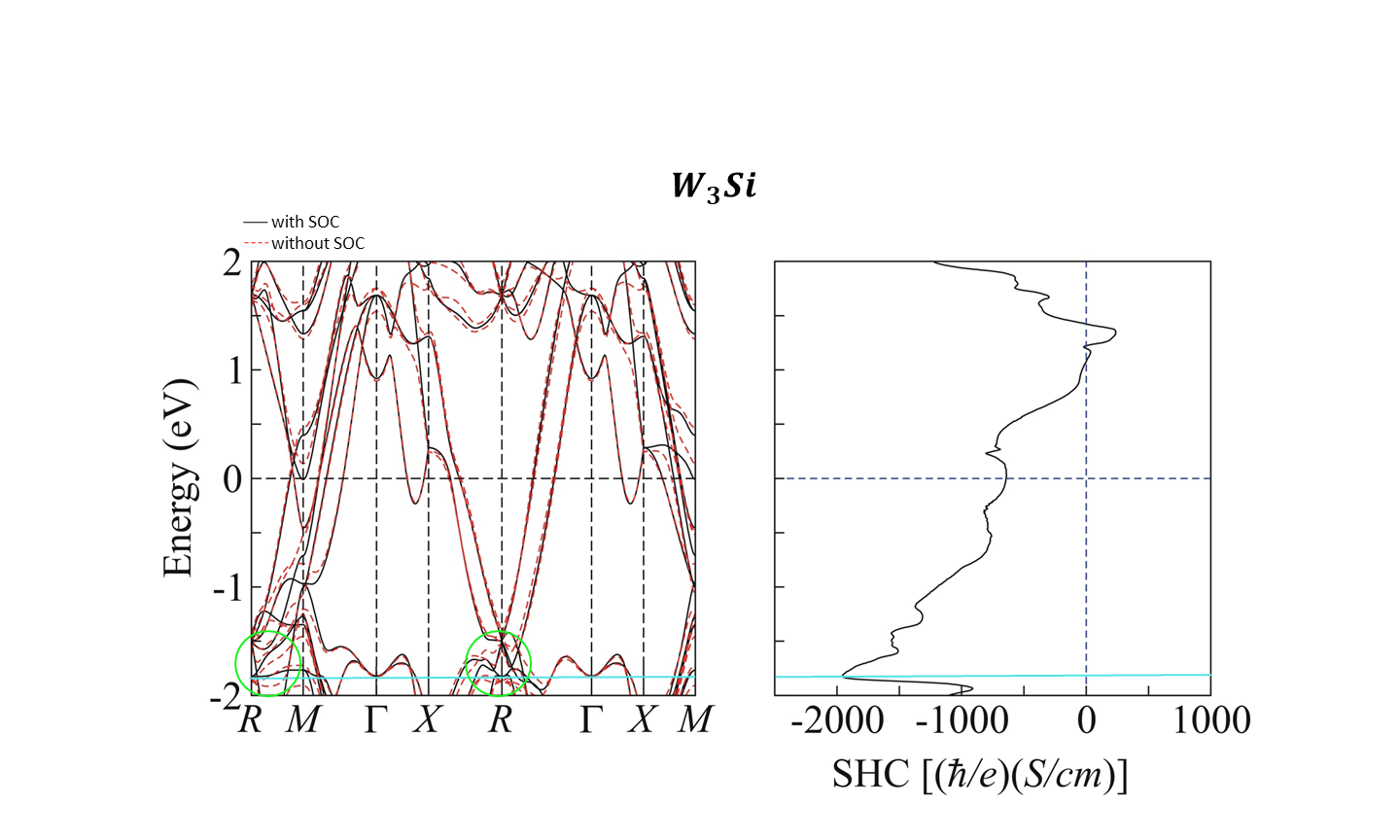}
	\caption{\scriptsize{\textbf{(color online):} Electronic structure without and with SOC included as well as the SHC vs energy plot for W$_3$Si}}
	\label{Figure_S3}
\end{figure}


\begin{figure}[ht]
	\includegraphics[width=0.8\textwidth]{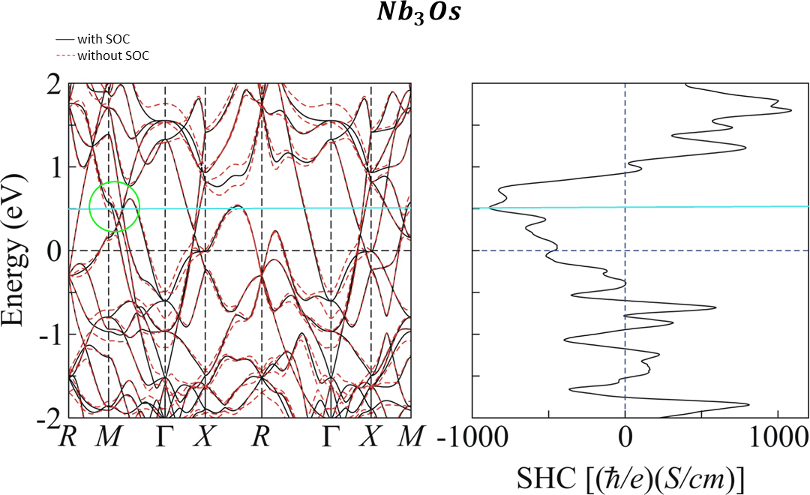}
 \caption{\scriptsize{\textbf{(color online):} Electronic structure without and with SOC included as well as the SHC vs energy plot for Nb$_3$Os}}
	\label{Figure_S4}
\end{figure}


\begin{figure}[ht]
	\includegraphics[width=1\textwidth]{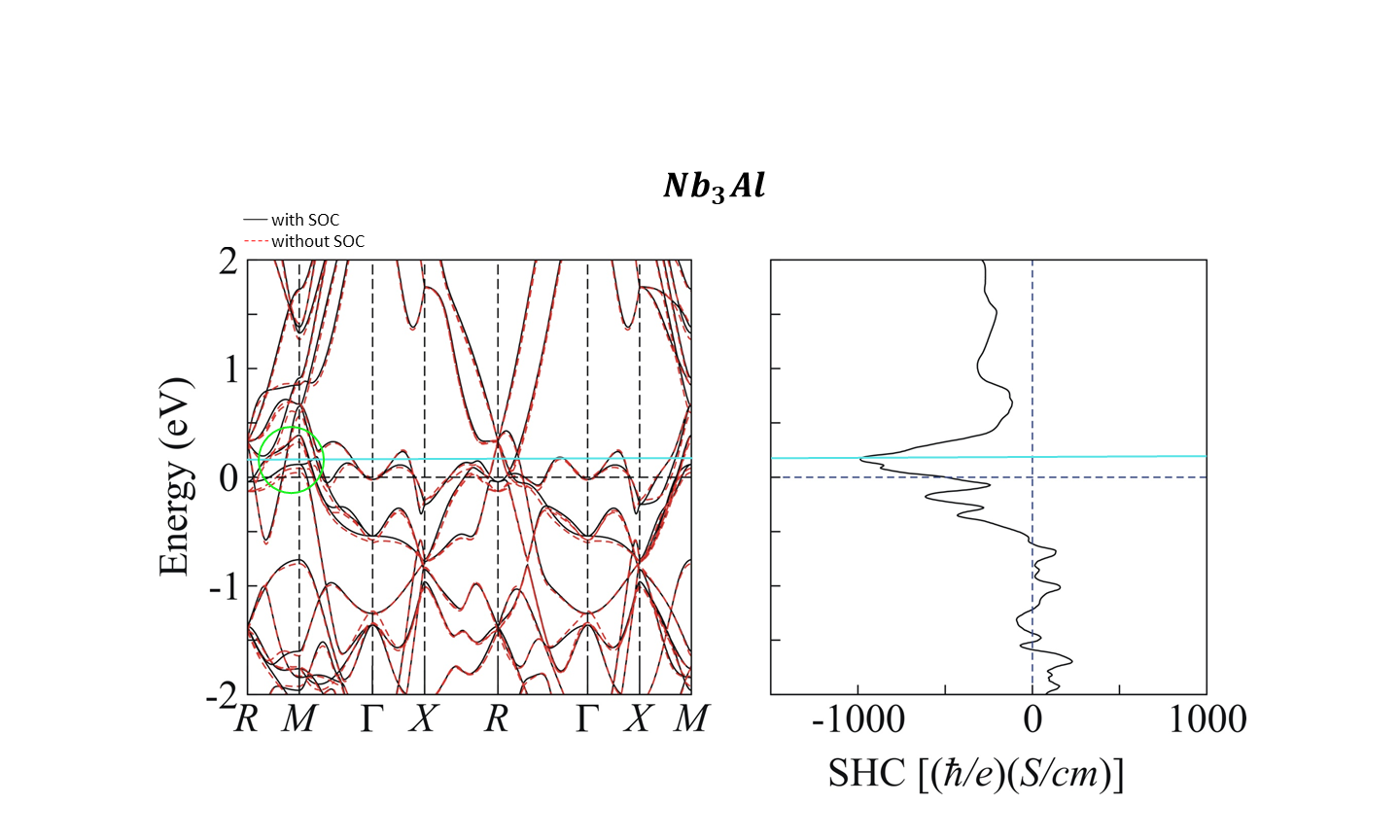}
	\caption{\scriptsize{\textbf{(color online):} Electronic structure without and with SOC included as well as the SHC vs energy plot for Nb$_3$Al}}
	\label{Figure_S5}
\end{figure}


\begin{figure}[ht]
	\includegraphics[width=1\textwidth]{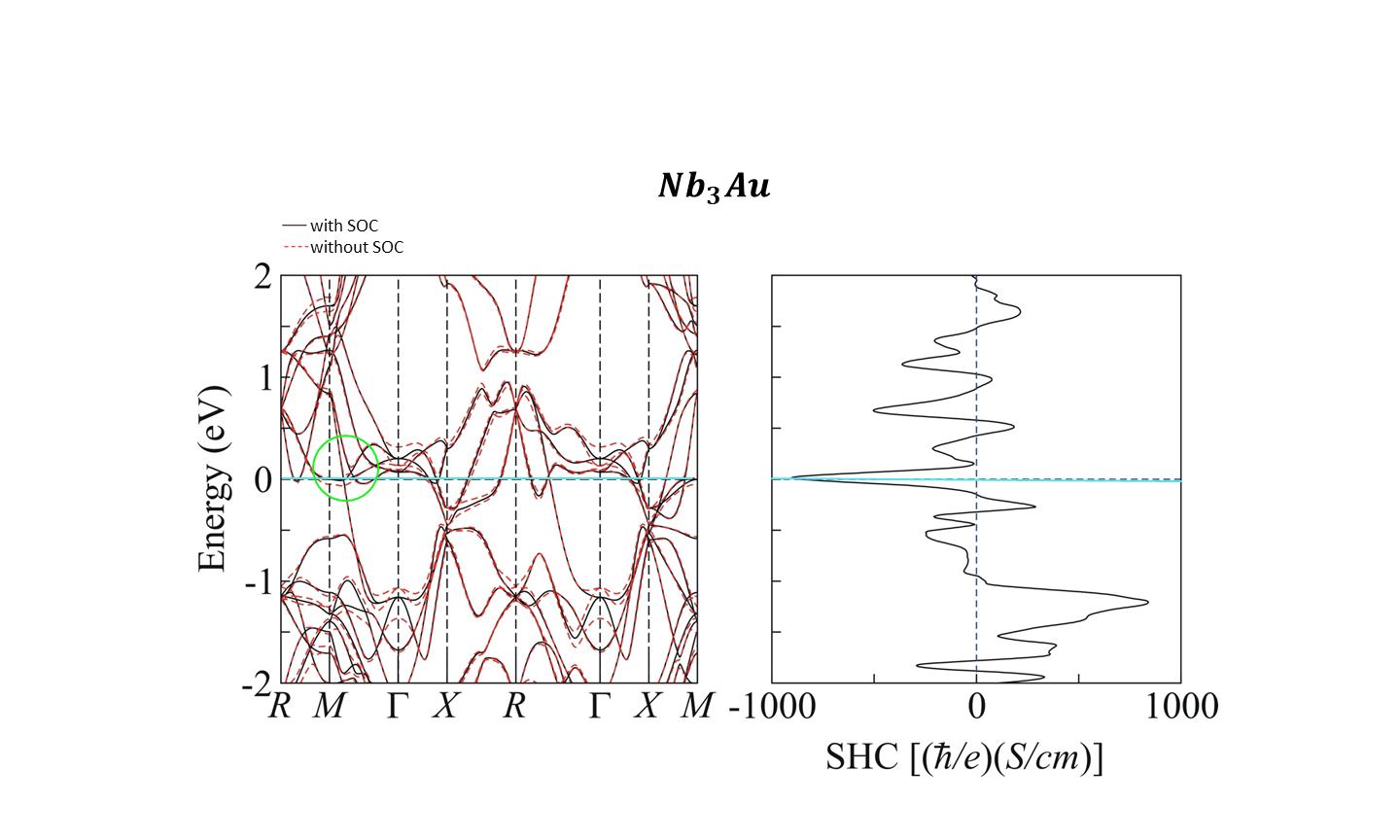}
	\caption{\scriptsize{\textbf{(color online):} Electronic structure without and with SOC included as well as the SHC vs energy plot for Nb$_3$Au}}
	\label{Figure_S6}
\end{figure} 


\begin{figure}[ht]
	\includegraphics[width=1\textwidth]{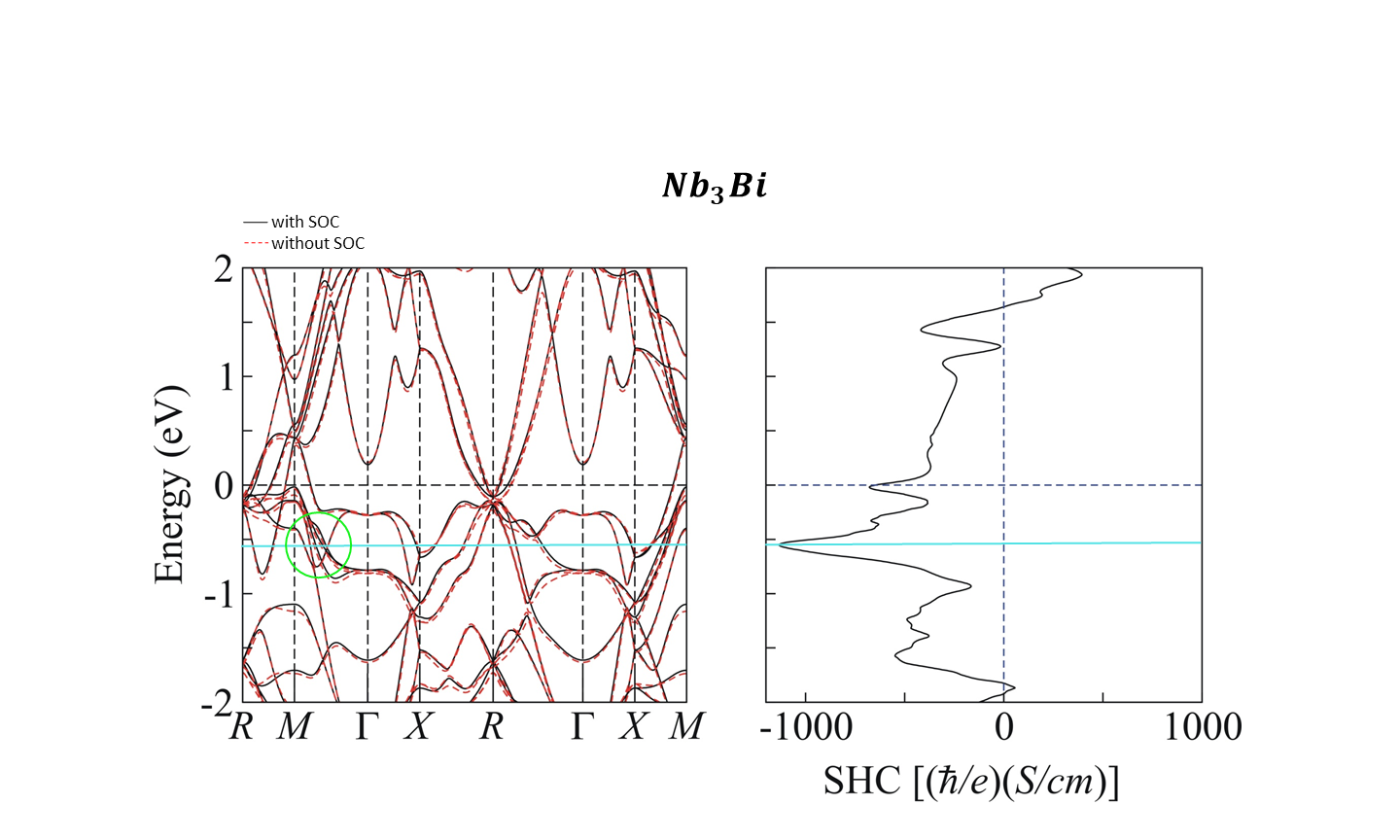}
	\caption{\scriptsize{\textbf{(color online):} Electronic structure without and with SOC included as well as the SHC vs energy plot for Nb$_3$Bi}}
	\label{Figure_S7}
\end{figure}


\begin{figure}[ht]
	\includegraphics[width=1\textwidth]{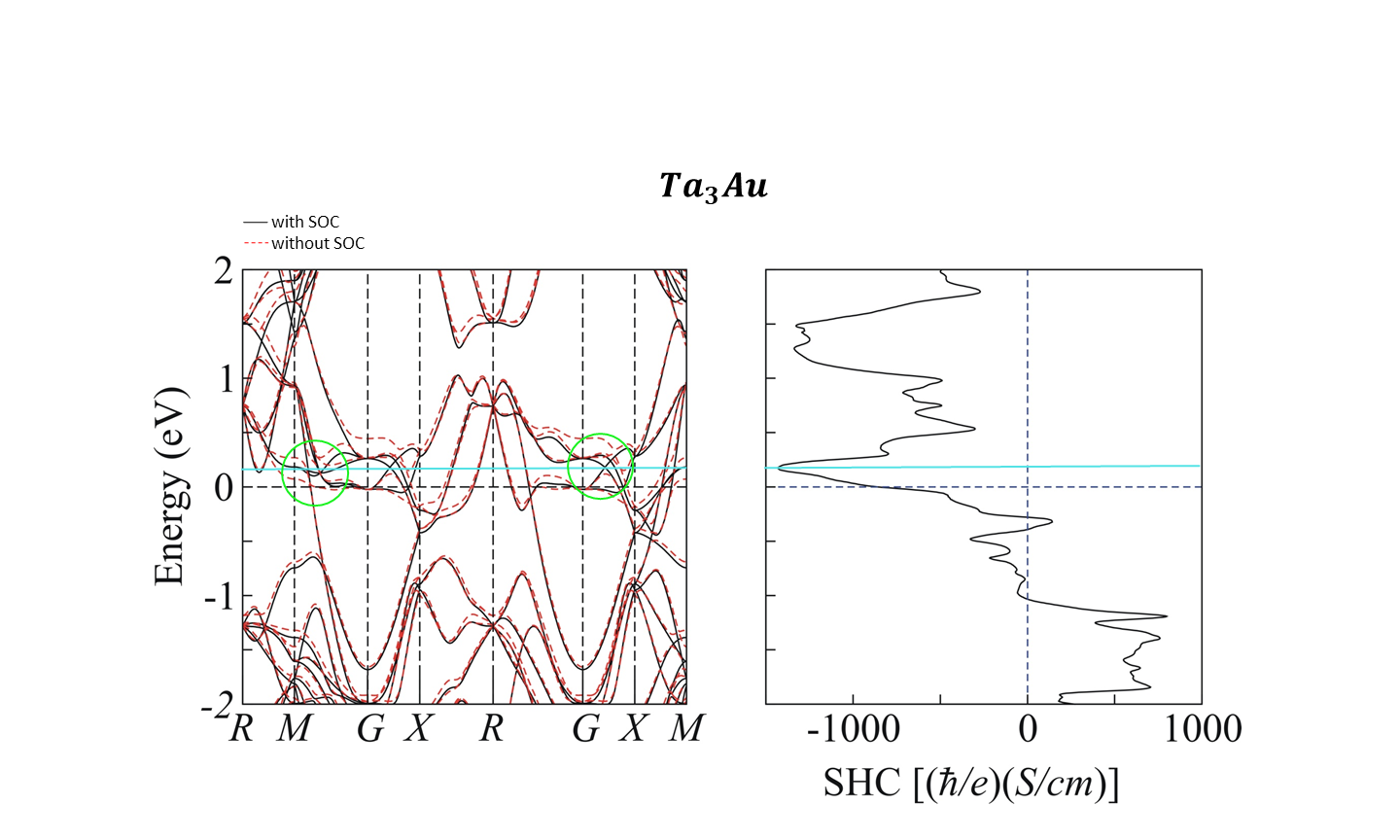}
 \caption{\scriptsize{\textbf{(color online):} Electronic structure without and with SOC included as well as the SHC vs energy plot for Ta$_3$Au}}
	\label{Figure_S8}
\end{figure}


\begin{figure}[ht]
	\includegraphics[width=0.8\textwidth]{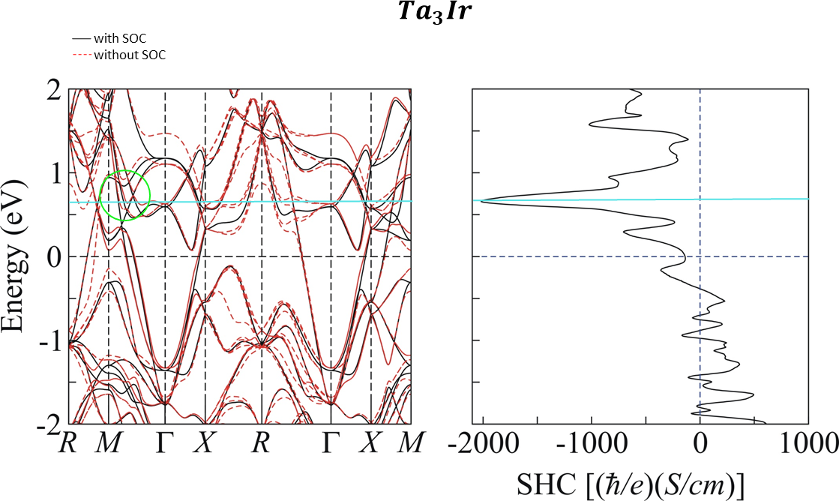}
	\caption{\scriptsize{\textbf{(color online):} Electronic structure without and with SOC included as well as the SHC vs energy plot for Ta$_3$Ir}}
	\label{Figure_S9}
\end{figure}


\begin{figure}[ht]
	\includegraphics[width=0.8\textwidth]{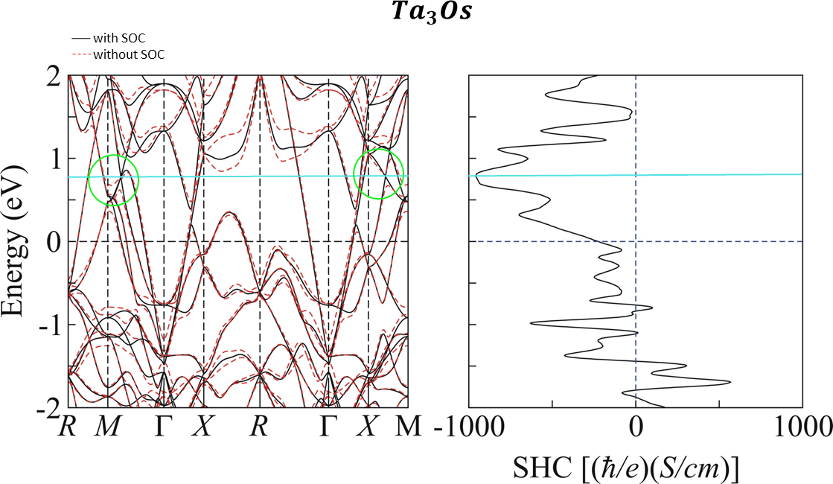}
	\caption{\scriptsize{\textbf{(color online):} Electronic structure without and with SOC included as well as the SHC vs energy plot for Ta$_3$Os}}
	\label{Figure_S10}
\end{figure}


\begin{figure}[ht]
	\includegraphics[width=1\textwidth]{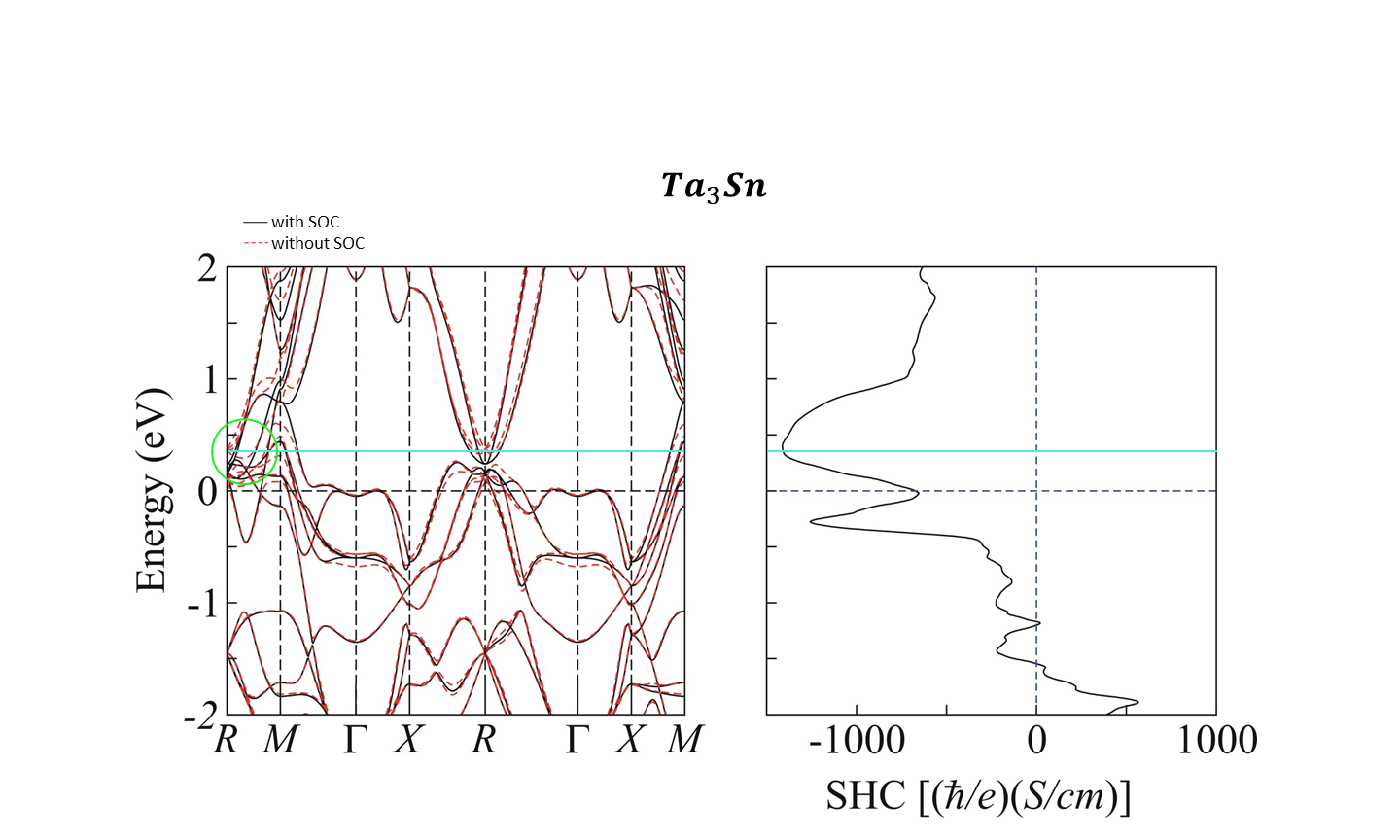}
	\caption{\scriptsize{\textbf{(color online):} Electronic structure without and with SOC included as well as the SHC vs energy plot for Ta$_3$Sn}}
	\label{Figure_S11}
\end{figure} 


\begin{figure}[ht]
	\includegraphics[width=0.8\textwidth]{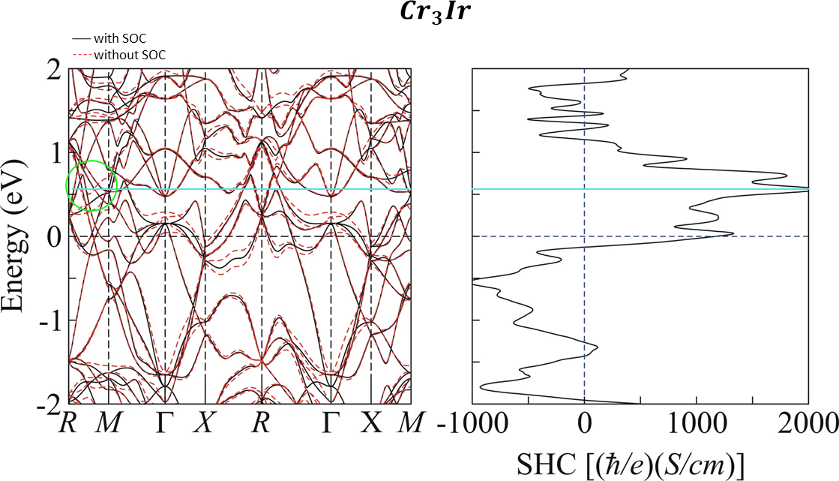}
	\caption{\scriptsize{\textbf{(color online):} Electronic structure without and with SOC included as well as the SHC vs energy plot for Cr$_3$Ir}}
	\label{Figure_S12}
\end{figure}


\begin{figure}[ht]
	\includegraphics[width=0.8\textwidth]{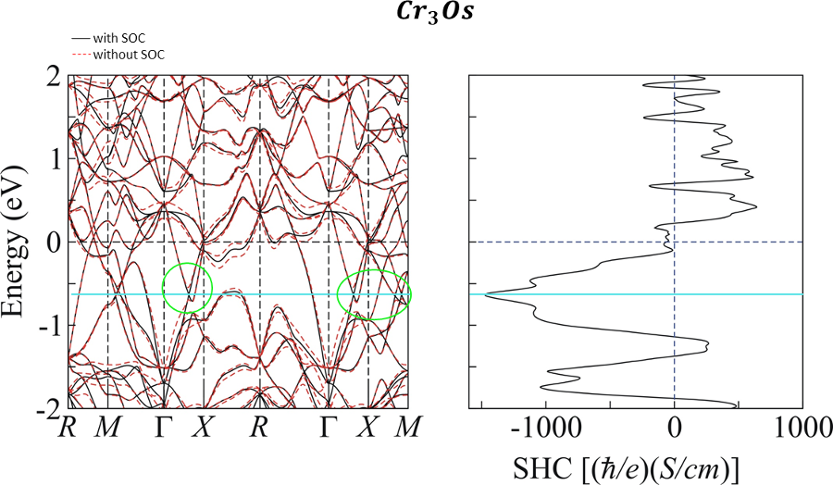}
 \caption{\scriptsize{\textbf{(color online):} Electronic structure without and with SOC included as well as the SHC vs energy plot for Cr$_3$Os}}
	\label{Figure_S13}
\end{figure}


\begin{figure}[ht]
	\includegraphics[width=1\textwidth]{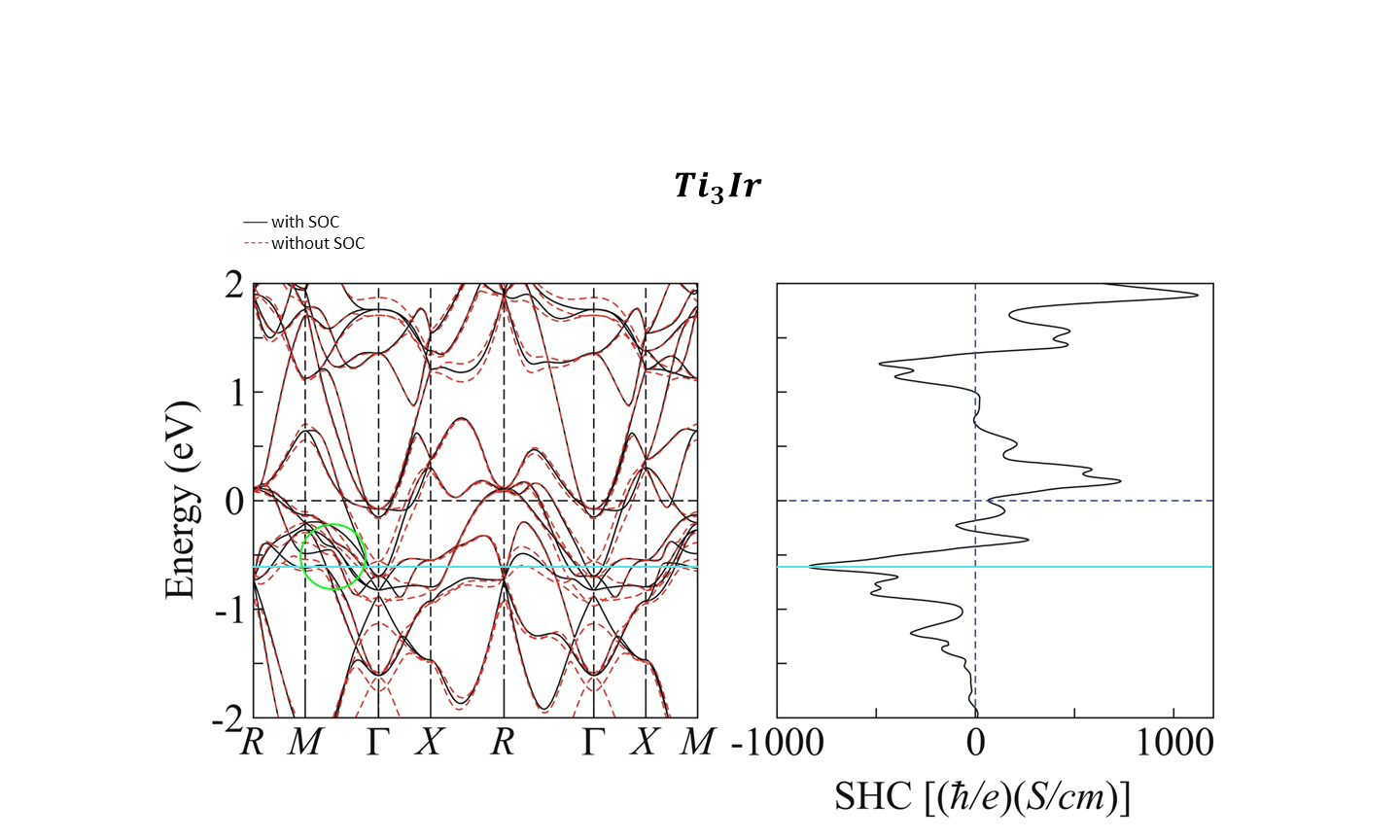}
	\caption{\scriptsize{\textbf{(color online):} Electronic structure without and with SOC included as well as the SHC vs energy plot for Ti$_3$Ir}}
	\label{Figure_S14}
\end{figure}


\begin{figure}[ht]
	\includegraphics[width=1\textwidth]{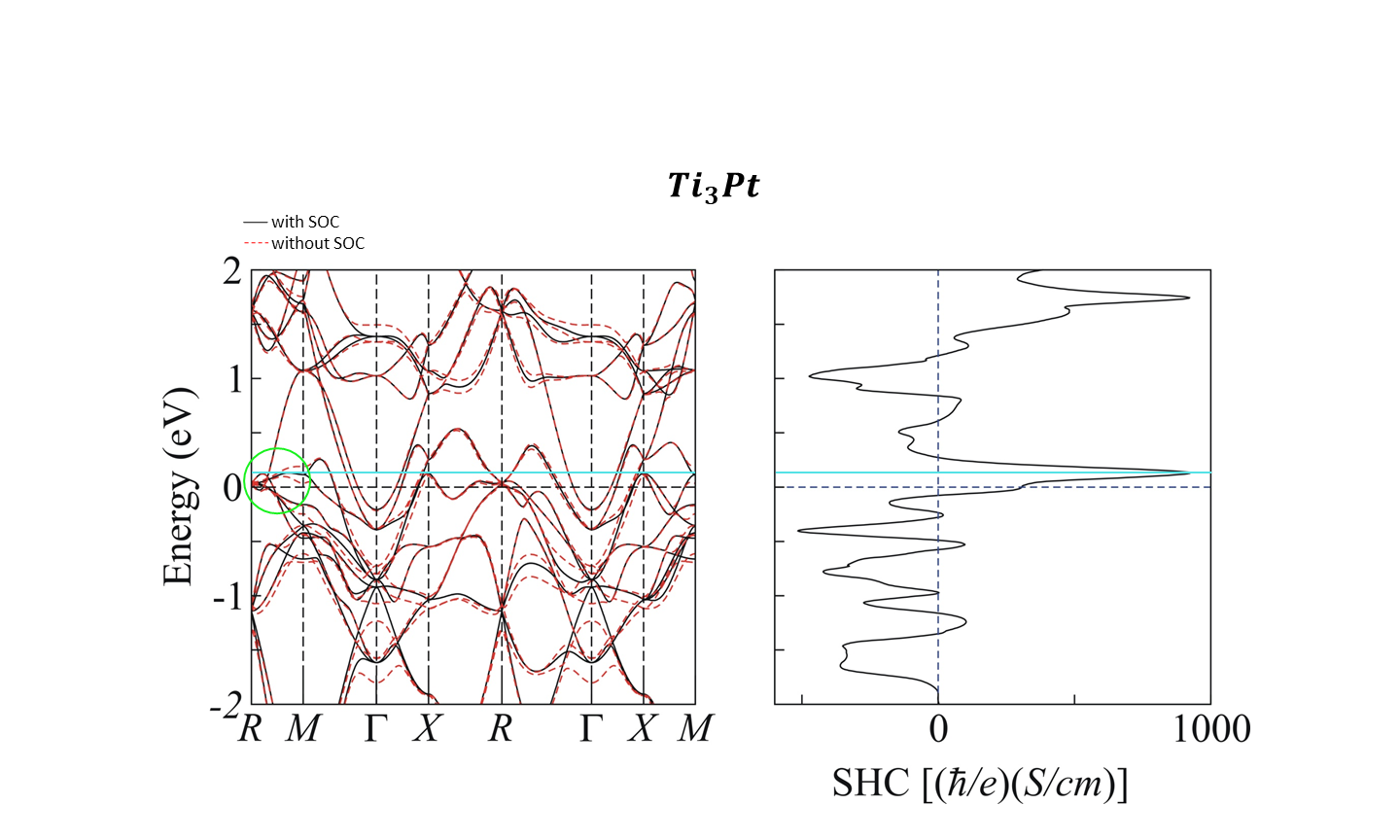}
	\caption{\scriptsize{\textbf{(color online):} Electronic structure without and with SOC included as well as the SHC vs energy plot for Ti$_3$Pt}}
	\label{Figure_S15}
\end{figure} 


\begin{figure}[ht]
	\includegraphics[width=0.8\textwidth]{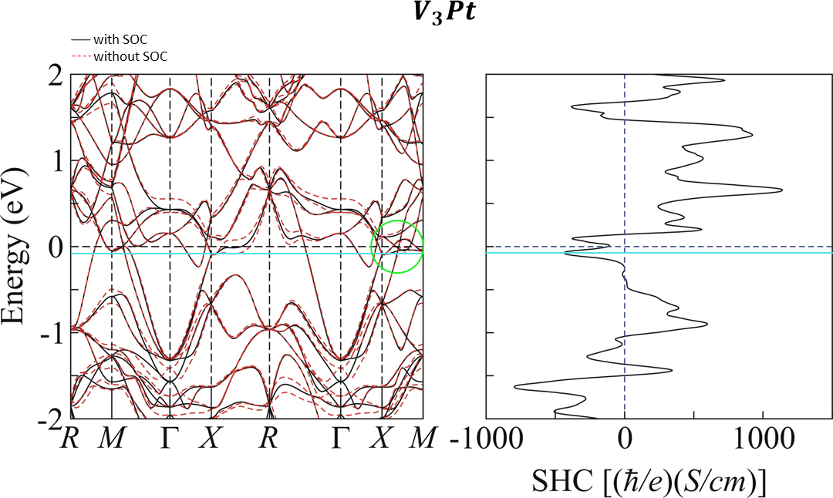}
	\caption{\scriptsize{\textbf{(color online):} Electronic structure without and with SOC included as well as the SHC vs energy plot for V$_3$Pt}}
	\label{Figure_S16}
\end{figure} 

\end{document}